\documentclass[preprint,nofootinbib,superscriptaddress]{revtex4}

\usepackage{amssymb}
\usepackage{amsmath} 
\usepackage{mathrsfs}
\usepackage{bbold}
\usepackage{tensor}
\usepackage{epsfig}
\usepackage{pstricks}
\usepackage{pst-node}
\usepackage{feynmp}

\def\be{\begin{equation}}
\def\ee{\end{equation}}
\def\ba{\begin{eqnarray}}
\def\ea{\end{eqnarray}}	
\def\l{\left}
\def\r{\right}
\def\fr{\frac}
\def\la{\label}
\def\d{\partial}
\def\vphi{\varphi}

\begin{document}

\begin{fmffile}{graphs}



\title{Three Approaches to Classical Thermal Field Theory}




\author{E.Gozzi}
\affiliation{Department of  Physics, University of Trieste, Strada Costiera 11, Miramare - Grignano, 34014 Trieste, Italy}
\affiliation{INFN, Sezione di Trieste, Italy}

\author{R.Penco}
\affiliation{Department of Physics, Syracuse University, Syracuse, NY 13244-1130, USA}

\begin{abstract}
In this paper we study three different functional approaches to {\it classical} thermal field theory, which turn out to be the classical counterparts of three well-known different formulations of {\it quantum} thermal field theory: the Closed-Time Path (CTP) formalism, the Thermofield Dynamics (TFD) and the Matsubara approach. 
\end{abstract}


\maketitle



\tableofcontents

\section{Introduction}

In recent years, there has been a growing interest in the physics of collisions between heavy ions \cite{d'Enterria:2006cg, Tannenbaum:2007dx}. In particular, one of the most interesting topics in the field has become the study of the quark-gluon plasma (QGP), which is a state of baryonic matter characterized by high values of temperature and density and by the fact that quarks are not confined \cite{QGP}. A better understanding of the physics of this state would be an important step forward in high energy physics and it could have important implications for early universe cosmology as well \cite{Grigoriev:1988bd}.

The theoretical framework which is required to study the QGP is the finite temperature quantum field theory. In fact, this formalism can take into account the quantum, the relativistic and the statistical features of the system at the same time. There are evidences that in the QGP gluons are characterized by very large occupation numbers. In this state, one would expect the system to be described by a classical field theory \cite{Mueller:2002gd}. In fact, it is known that QED is well-approximated by the classical Maxwell theory when occupation numbers of photons become large. Analogously, when occupation numbers of gluons are large, it should be possible to approximate QCD with a classical gauge field theory \cite{Stockamp:2004qu}.

These considerations have led various authors to study classical field theories at high temperatures and, to our knowledge, three main approaches have been pursued in the literature so far. The first approach is due to Aarts and collaborators \cite{Aarts:1996qi, Aarts:1997kp, Aarts:1997ve, Aarts:1999wj, Aarts:2001yx}, who computed classical thermal correlation functions $\langle \phi(x) \cdots \phi(x^{\prime})\rangle_{\beta}$ by solving the Hamiltonian equations of motion and averaging the solutions over a canonical ensemble of initial conditions. An alternative approach has been proposed by F. Cooper {\it et al.} \cite{Cooper:2001bd} and it is based on the operatorial formalism developed by P.C. Martin, E.D. Siggia and H.A. Rose \cite{MSR} in the 1970s and known as MSR formalism. Finally, S.Jeon \cite{Jeon:2004dh} has recently studied a classical scalar field theory at finite temperature by resorting to the path integral approach to classical mechanics proposed in ref.  \cite{Gozzi:1989bf} at the end of the 80s.

Despite the formal differences, there is a common feature shared by the three classical approaches mentioned above: they are all related to the quantum Closed Time Path (CTP) formalism \cite{Schwinger:1960qe, Keldysh:1964ud} in the high temperature limit. However, quantum field theory at finite temperature admits several other equivalent formulations, the most popular ones being the Matsubara formalism \cite{Matsubara:1955ws} and the Thermofield Dynamics (TFD) \cite{Umezawa:1982nv}. The goal of this paper is to study the ``classical analogs'' of the TFD and the Matsubara formalism as well. To this end, we will present a general framework for classical thermal field theory based on the operatorial approach to classical mechanics  proposed in the 30's by Koopman  and von Neumann (KvN) \cite{Koopman, vonNeumann1, vonNeumann2} and on the associated classical path-integral (CPI) formulation \cite{Gozzi:1989bf}. We believe that these tools (KvN and CPI) are the most suitable ones for studying the interplay between classical and quantum mechanics \cite{Abrikosov:2004cf} either with or without temperature. Within this theoretical framework, we will show how the {\it classical} finite temperature field theory can be formulated in three different ways, which are in a certain sense the ``{\it classical} counterparts" of the three {\it quantum} formalisms mentioned above: the CTP, the TFD and the Matsubara one. All these formalisms will be illustrated by considering the simple example of a scalar field with quartic self-interaction. 

The paper is organized as follows: after reviewing briefly the various equivalent approaches to quantum thermal field theory (Section \ref{sec:QTFT}) and the KvN and CPI formulations of classical mechanics (Section \ref{sec:CPIKvN}), in Section \ref{sec:classicalCTP} we will present the approach to classical thermal field theory developed in \cite{Jeon:2004dh}. From a physical point of view, this is the most intuitive approach to classical thermal field theory, as it is based on the same principles used by Aarts and Smit in \cite{Aarts:1996qi, Aarts:1997kp}, i.e. on solving the equations of motion and averaging over the initial conditions using the canonical distribution. As we already mentioned, this approach turns out to be related to the CTP approach to quantum thermal field theory in the high temperature limit.

Then, in the second part of the paper we shall develop two new approaches to {\it classical} thermal field theory which display both a remarkable formal analogy and a quantitative agreement with their {\it quantum} counterparts at high temperatures, i.e.  the TFD approach and the Matsubara formalism. In particular, in Section \ref{sec:classicalTFD} we will study the TFD approach to {\it classical} thermal field theory, which is based on the idea of implementing thermal averages as expectation values on a particular state $|\psi_{\beta}\rangle$ in the {\it classical} Hilbert space of ref. \cite{ Deotto:2002hy}. In Section \ref{sec:classicalMatsubara} we will study the classical counterpart of the quantum Matsubara formalism. Both quantum and classical Matsubara approaches allow us to compute only static properties of system, since the time variable is formally restricted to the purely imaginary axis. However, while in the quantum case time is an element of a complex plane, in the classical case time is an element of a complex {\it  superspace} \cite{DeWitt:1992cy}. As we will see, the appearance of superspace is just a natural consequence of the fact that the path integral formulation of classical mechanics features a universal $N=2$ supersymmetry and, as such, admits a very compact representation in terms of some suitably defined superfields \cite{Abrikosov:2004cf}. Finally, some further background material as well as some technical details were included in two appendices.

\section{Quantum thermal field theory} \la{sec:QTFT}

In order to make the paper as self-contained as possible, in this section we will review some aspects of quantum field theory at finite temperature. For a more exhaustive treatment we refer the reader to the review articles \cite{Niemi:1983nf, Altherr:1993tn} or the textbooks \cite{Das:1997gg, LeBellac:1996}. The main goal of this section is to show how perturbative calculations can be implemented in several different ways, and to derive the corresponding sets of Feynman rules. For concreteness, the main concepts will be illustrated by considering the simple case of a scalar field described by the Hamiltonian
\begin{equation} \la{hamiltonian}
H = \int d^3 x \left\{ \frac{\pi^2}{2} + \frac{(\nabla \phi)^2}{2} + \frac{m^2 \phi^2}{2}  + \fr{g \, \phi^4}{4!} \right\}, 
\end{equation}
where $\pi$ is the canonical momentum associated with $\phi$.

In the case of a scalar field in thermal equilibrium at temperature $T=1/\beta$, the quantities of interest are the thermal correlation functions, which are defined as expectation values of time-ordered products of fields $\phi$:
\be \la{thermalcorrelationfunctions}
G_\beta (x_1, ..., x_n) \equiv \langle T [ \phi(x_1) \cdots \phi(x_n) ] \rangle_\beta = \fr{\mbox{Tr} \l\{ e^{- \beta \hat{H}} \, T [ \hat{\phi}(x_1) \cdots \hat{\phi}(x_n) ] \r\} }{\mbox{Tr} \,e^{- \beta \hat{H}}},
\ee
where $T[\cdots]$ stands for the time-ordered product and $\hat{\rho} = e^{- \beta \hat{H}}$ is the statistical operator describing a canonical ensemble. These quantities are clearly a generalization of the correlation functions one encounters in ordinary quantum field theory, since they reduce to the vacuum expectation value of a time-ordered product of fields in the zero-temperature limit $\beta \to \infty$:
\begin{equation}
 G_{\beta} (x_1, ..., x_n) =  \fr{\sum_n e^{- \beta E_n} \langle n | \, T [ \hat{\phi}(x_1) \cdots \hat{\phi}(x_n) ] | n \rangle}{\sum_n e^{- \beta E_n}} \quad \stackrel{\beta \to \infty}{\longrightarrow} \quad \langle 0 | \, T [ \hat{\phi}(x_1) \cdots \hat{\phi}(x_n) ] | 0 \rangle.
\end{equation}

Before deriving a set of Feynman rules which will allow us to calculate thermal correlation functions in a systematic way, it is useful to derive an important result known as Kubo-Martin-Schwinger (KMS) relation \cite{Kubo:1957mj, Martin:1959jp}. This result is based on the simple idea that, if one allows time to be a complex variable, then the statistical operator $e^{- \beta \hat{H}}$ can be interpreted as a time evolution operator with a purely imaginary argument $- i \beta$. As a consequence, the thermal average of the product of any two observables $A$ and $B$ in the Heisenberg picture satisfies the following identity:
\be \la{KMS}
\langle A(t) B(t') \rangle_\beta = \fr{\mbox{Tr} [e^{- \beta \hat{H}} \hat{A}(t) \hat{B}(t') ]}{\mbox{Tr} \, e^{- \beta \hat{H}}} = \fr{\mbox{Tr} [e^{- \beta \hat{H}} \hat{B}(t')  \hat{A}(t+i \beta) ]}{\mbox{Tr} \, e^{- \beta \hat{H}}} = \langle  B(t') A(t+i \beta) \rangle_\beta .
\ee
This identity is the celebrated KMS relation. We should point out that this relation was obtained by using the cyclicity of the trace, and this property may fail if the trace is not finite \cite{Das:1997gg}. 
We will encounter an example of such a behavior in Section \ref{sec:classicalMatsubara}.

\subsection{Path Integral Formulation}

The idea of interpreting the statistical operator $e^{- \beta \hat{H}}$ as a time evolution operator not only allow us to derive an important identity such as the KMS relation, but it also leads naturally to a path integral expression for the canonical partition function $Z_\beta = \mbox{Tr} \, e^{- \beta \hat{H}}$ \cite{Bernard:74,Jackiw:74}. In fact, by expanding the trace on the basis of the eigenstates of the field operator $\hat{\phi}(t, \mathbf{x})$ at a give time $t_0$, we get
\be
Z_\beta  = \int [d \phi] \langle \phi, t_0 | e^{- \beta \hat{H}} | \phi, t_0 \rangle = \int [d \phi]  \langle \phi, t_0 - i \beta | \phi, t_0 \rangle,  \la{transampl}
\ee
where $[d \phi]$ denotes a path integral over all the field configurations $\phi(t_0, \mathbf{x})$ with fixed time $t_0$. By resorting to the standard slicing procedure, the transition amplitude on the RHS of equation (\ref{transampl}) can be easily expressed in terms of a path integral over all the field configurations  that satisfy the periodic boundary condition $\phi(t_0, \mathbf{x}) = \phi(t_0-i\beta, \mathbf{x})$. In the particular case of a scalar field described by the Hamiltonian (\ref{hamiltonian}), we get
\be \la{partfuncpathintexpr}
Z_\beta = \int \mathcal{D}_{\scriptscriptstyle{\mathcal{C}}} \, \phi \, e^{i S}, \qquad \quad S = \int_\mathcal{C} dt \int d^3x \l\{ \fr{1}{2} \d_\mu \phi \d^\mu \phi - \fr{m^2}{2} \phi^2 - \fr{g}{4!} \phi^4 \r\},
\ee
where $\mathcal{C}$ is a path in the complex plane of time which starts at $t_0$ and ends at $t_0-i\beta$. Notice that the time $t_0$ is arbitrary, and all that matters is that the initial and final points of the time contour $\mathcal{C}$ are separated by an amount $- i \beta$. The path integral expression (\ref{partfuncpathintexpr}) immediately suggests that thermal correlation functions evaluated at times along the path $\mathcal{C}$ could be systematically calculated starting from the following generating functional:
\be \la{generalgeneratingfunctional}
Z[j] = \fr{\displaystyle \int \mathcal{D}_{\scriptscriptstyle{\mathcal{C}}} \, \phi \,  \exp \l\{ i S + i \int_\mathcal{C} dt \int d^3 x \, j \phi \r\}}{\displaystyle \int \mathcal{D}_{\scriptscriptstyle{\mathcal{C}}} \, \phi \, \exp \l\{ i S \r\} }.
\ee
However, consistency requires that the time contour $\mathcal{C}$ lies within the domain of analyticity of the thermal correlation functions $G_\beta (x_1, \cdots, x_n)$. This constraint is satisfied provided the time path $\mathcal{C}$ is nowhere directed upward in the complex plane of time \cite{Niemi:1983nf}. Notice also that the time ordered product that appears in equation (\ref{thermalcorrelationfunctions}) refers now to the order along the time path $\mathcal{C}$. 

Starting from equation (\ref{generalgeneratingfunctional}), a set of Feynman rules can be derived as usual by treating the quartic coupling in (\ref{partfuncpathintexpr}) as a perturbation. In particular, the propagator can be determined by considering just the quadratic terms in equation (\ref{partfuncpathintexpr}). In this case, the generating functional can be calculated exactly and takes the form
\begin{equation} \label{generalfreegeneratingfunctional}
Z_0 [ j ] = \exp \left\{ - \frac{1}{2} \int_\mathcal{C} d^4 x \, d^4 x^{\prime} \, j(x) G_{\beta} (x-x^{\prime})  j(x^{\prime}) \right\},
\end{equation}
where $G_{\beta} (x-x^{\prime})$ is the propagator, i.e. the thermal correlation function of two fields $\phi$, which satisfies the following differential equation:
\begin{equation} \la{freeequation}
\left( \square_{\, \scriptscriptstyle{\mathcal{C}}} + m^2 \right)  G_{\beta} (x-x^{\prime}) = - i \, \delta_{\scriptscriptstyle{\mathcal{C}}} (t - t^{\prime}) \delta ({\bf x} - {\bf x}^{\prime}).
\end{equation}
Notice that, in this equation, both the differential operator $\square_{\, \scriptscriptstyle{\mathcal{C}}}$ and the Dirac delta $\delta_{ \scriptscriptstyle{\mathcal{C}}} (t - t^{\prime})$ are defined along the time contour $\mathcal{C}$. According to the definition (\ref{thermalcorrelationfunctions}) of thermal correlation functions, the propagator $G_{\beta} (x-x^{\prime})$ can be written as
\begin{equation} 
G_{\beta} (x-x^{\prime}) = G_{>} (x-x^{\prime}) \, \theta_{ \scriptscriptstyle{\mathcal{C}}} (t - t^{\prime}) +  G_{<} (x-x^{\prime}) \, \theta_{ \scriptscriptstyle{\mathcal{C}}} (t^{\prime} - t),
\end{equation}
where  $G_{>} (x-x^{\prime}) \equiv  \langle \phi (x) \phi(x^{\prime})\rangle_{\beta}$,  $G_{<} (x-x^{\prime}) \equiv \langle \phi (x^{\prime}) \phi(x)\rangle_{\beta}$ and the Heaviside step function along the contour $\mathcal{C}$ is related to the Dirac delta in equation (\ref{freeequation}) by $\fr{d}{dt} \theta_{ \scriptscriptstyle{\mathcal{C}}} (t - t^{\prime}) = \delta_{ \scriptscriptstyle{\mathcal{C}}} (t - t^{\prime})$. The two functions $G_>$ and $G_<$ are not independent of each other, since the KMS relation (\ref{KMS}) implies that
\begin{equation}\label{boundaryKMS}
G_{>} (t-t^{\prime}, \mathbf{x}-\mathbf{x}^{\prime}) = G_{<} (t-t^{\prime} + i \beta, \mathbf{x}-\mathbf{x}^{\prime}).
\end{equation}
The propagator $G_{\beta} (x-x^{\prime})$ can then be determined univocally by solving the differential equation (\ref{freeequation}) together with the boundary condition (\ref{boundaryKMS}). After performing a Fourier transformation with respect to the spatial variables only, the propagator in the $(t, \mathbf{p})$ space reads~\cite{Niemi:1983nf}
\begin{eqnarray} \la{propagator}
\parbox{20mm}{
\begin{fmfgraph}(65,30)
\fmfleft{i1}
\fmfright{o1}
\fmf{phantom}{i1,o1}
\fmffreeze
\fmf{vanilla}{i1,o1}
\fmfv{decor.shape=circle, decor.filled=full, decor.size=1.5thick}{i1,o1}
\end{fmfgraph}} \quad = G_{\beta} (t - t^{\prime}, {\bf p}) &=& \frac{n_B( E_{{\bf p}})}{2 E_{{\bf p}}} \left[ \left( e^{ i E_{{\bf p}} (t - t^{\prime})}  + e^{\beta E_{{\bf p}}} e^{- i E_{{\bf p}} (t - t^{\prime})} \right) \theta_{ \scriptscriptstyle{\mathcal{C}}} (t - t^{\prime}) \, + \right.\nonumber \\
&& \qquad \qquad + \left. \left( e^{ - i E_{{\bf p}} (t - t^{\prime})}  + e^{\beta E_{{\bf p}}} e^{ i E_{{\bf p}} (t - t^{\prime})} \right) \theta_{ \scriptscriptstyle{\mathcal{C}}} (t^{\prime} - t) \right] \label{h.2}
\end{eqnarray}
where we defined $E_{{\bf p}}^2 \equiv  {\bf p}^2 + m^2$, while $n_B (E_{{\bf p}}) \equiv (e^{\beta E_{{\bf p}}} - 1)^{-1}$  is the Bose-Einstein density of states. The effect of the quartic coupling can then be taken into account as usual by introducing a vertex with the following rule in the $(t, \mathbf{p})$ space:
\begin{equation} \label{vertex}
\parbox{20mm}{
\begin{fmfgraph}(45,40)
\fmfleft{i1,i2}
\fmfright{o1,o2}
\fmf{vanilla}{i1,v,o2}
\fmf{vanilla}{i2,v,o1}
\fmfv{decor.shape=circle, decor.filled=full, decor.size=1.5thick}{v}
\end{fmfgraph}}
\! \! \! \! = - i g \int_\mathcal{C} dt .
\end{equation}

Although the Feynman rules (\ref{propagator}) and (\ref{vertex}) apply to any path $\mathcal{C}$ which is nowhere directed upward in the complex plane of time, concrete applications usually become simpler once a specific contour is chosen. In the remaining part of this section we will briefly review some of the most popular choices.

\subsection{Matsubara Formalism}

\begin{figure}[b]
\begin{center}
\includegraphics[scale=0.6]{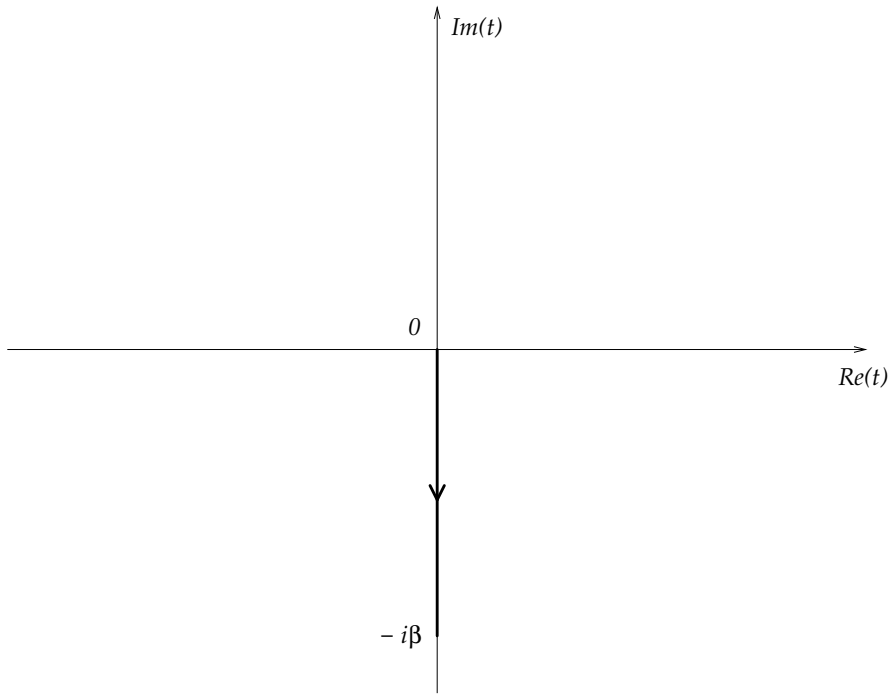}
\caption{Purely imaginary time path associated with the Matsubara formalism.}
\label{figA}
\end{center}
\end{figure}

The first approach to quantum field theory at finite temperature was originally developed by Matsubara \cite{Matsubara:1955ws} in the 50's  and it is based on a path the runs along the imaginary axis from $t=0$ to $t=-i\beta$, as shown in Figure \ref{figA}. Since we are ultimately interested in calculating thermal correlation functions at real times and the only intersection between the time contour in Figure \ref{figA} and the real axis is the starting point $t=0$, this formalism will only allow us to calculate time-independent correlation functions of fields $\phi(0,\mathbf{x})$. As such, the Matsubara formalism is a good tool for studying static properties of a system at equilibrium, but it is not suited to the study of time-dependent phenomena.

The path in Figure \ref{figA} can be conveniently parametrized by introducing the Euclidean time $t_E$ such that $t=-i t_E$. The main advantage of working with the Euclidean time as opposed to the time variable $t$ is that $t_E$ now belongs to the finite interval $[0, \beta]$, and since a trivial consequence of the KMS relation is that $G_\beta (-i\beta, \mathbf{p}) = G_\beta (0,\mathbf{p})$, the propagator $G_\beta (-i t_E, \mathbf{p})$ is a periodic function of $t_E$ on such interval. For this reason, it can be expressed as a Fourier series, and the Fourier coefficients can be easily calculated starting from the general expression (\ref{propagator}) for the propagator. The result one gets is
\be \la{Matsubarapropagator}
\parbox{20mm}{
\begin{fmfgraph}(65,30) 
\fmfleft{i1}
\fmfright{o1}
\fmf{phantom}{i1,o1}
\fmffreeze
\fmf{vanilla}{i1,o1}
\fmfv{decor.shape=circle, decor.filled=full, decor.size=1.5thick}{i1,o1}
\end{fmfgraph}} \quad = G_{\beta} (\omega_n, \mathbf{p}) = \int_0^\beta d t_E \, G_\beta (-i t_E, \mathbf{p}) \, e^{i \omega_n t_E} = \fr{1}{\omega_n^2 + E_{\mathbf{p}}^2} ,
\ee
where $n = 0,\pm1,\pm2, ...$ and the discrete frequencies $\omega_n = n \pi/\beta $ are known as \emph{Matsubara frequencies}. Furthermore, in the $(\omega_n, \mathbf{p})$ space, the rule (\ref{vertex}) for the vertex reduces to
\begin{equation} \label{Matsubaravertex}
\parbox{20mm}{
\begin{fmfgraph}(45,40)
\fmfleft{i1,i2}
\fmfright{o1,o2}
\fmf{vanilla}{i1,v,o2}
\fmf{vanilla}{i2,v,o1}
\fmfv{decor.shape=circle, decor.filled=full, decor.size=1.5thick}{v}
\end{fmfgraph}}
\! \! \! \! = - g .
\end{equation}

Clearly, one advantage of the Matsubara formalism is that its Feynman rules are fairly simple. In fact, for the simple model we are considering, they consist of only one propagator (see eq. (\ref{Matsubarapropagator})) and one vertex (see eq. (\ref{Matsubaravertex})). Notice also that these Feynman rules bear a close similarity to the ones encountered in ordinary quantum field theory, the only difference being that the discrete Matsubara frequencies have now replaced the continuous component $p^0$ of the four-momentum.  However, as we already mentioned, a severe limitation of this formalism is that it is not suitable for calculating dynamical properties of a system.

\subsection{Real Time Formalisms: CTP and Thermofield Dynamics}

\begin{figure}[t]
\begin{center}
\includegraphics[scale=0.6]{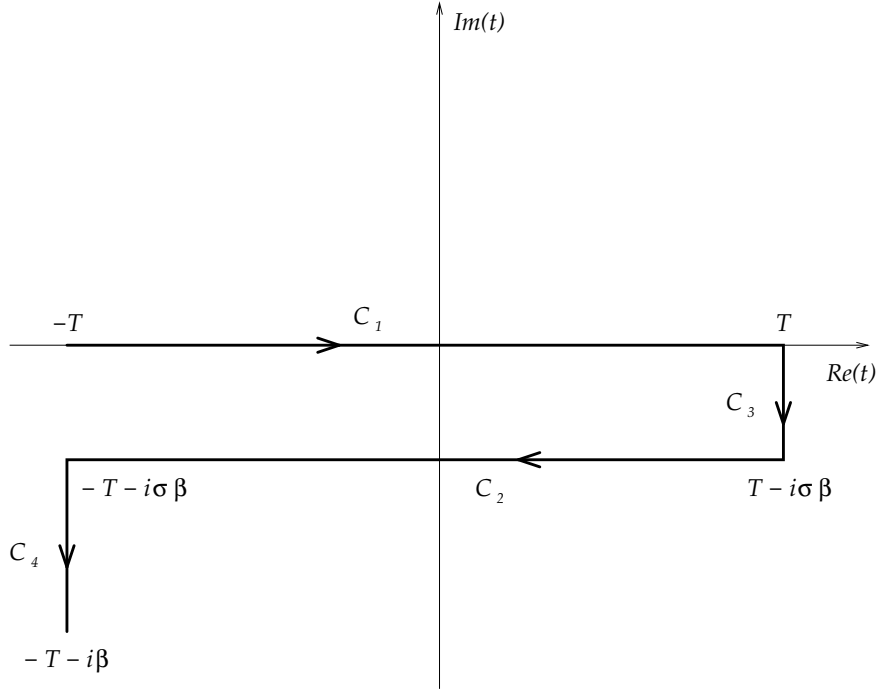}
\caption{Time path associated with real time formalisms.}
\label{figB}
\end{center}
\end{figure}

The main reason why the Matsubara formalism does not lend itself to the calculation of time-dependent properties of a system is that the time contour in Figure \ref{figA} does not lie along the real axis. In fact, dynamical properties can be calculated much more easily by working with the time path $\mathcal{C}$ shown in Figure \ref{figB}, which defines the so called \emph{real time formalisms}. Such path starts at $t=-T$, ends at $t = -T-i\beta$, and consists of four segments $\mathcal{C}_i$. The distance between the two horizontal segments $\mathcal{C}_1$ and $\mathcal{C}_2$ is controlled by the parameter $\sigma$. 
In the limit where $T \to \infty$, the two vertical segments $\mathcal{C}_3$ and $\mathcal{C}_4$ are pushed to infinity and therefore we expect their contribution to become negligible. This statement can be made more quantitative by breaking down the propagator $G_\beta$ into several different components according to the following definition:
\be \la{propagatorcomponents}
G_{i j} (t - t^{\prime}, \mathbf{p}) \equiv G_{\beta} (t - t^{\prime}, \mathbf{p}) \, e^{- \epsilon |Re(t-t')|} \qquad \mbox{with } \qquad t \in C_i, \, t^{ \prime} \in C_j.
\ee
So, for example, $G_{13}$ stands for the propagator that ``connects'' times on the segment $\mathcal{C}_1$ with times on the segment $\mathcal{C}_3$, and so on. Notice that, in equation (\ref{propagatorcomponents}), we have introduced an exponential term depending on the positive, infinitesimal parameter $\epsilon$. The introduction of this term is necessary to render the limit $T \to \infty$ well-behaved, and it is analogous to the ``$\epsilon$ convention'' usually adopted in ordinary quantum field theory. By combining the general expression (\ref{propagator}) for the propagator $G_\beta$ together with equation (\ref{propagatorcomponents}), it is easy to show that the propagators connecting times along horizontal segments with times along vertical segments vanish in the limit $T \to \infty$, i.e.
\be
G_{13}, \, G_{23}, \, G_{14}, \, G_{24} \quad \stackrel{T \to \infty}{\longrightarrow} \quad 0 .
\ee
Thus, horizontal and vertical segments effectively decouple in this limit and, since we are ultimately interested in calculating thermal correlation functions at times along the real axis (i.e. along the segment $\mathcal{C}_1$), the vertical segments $\mathcal{C}_3$ and $\mathcal{C}_4$ can be neglected.

At the same time, in the limit $T \to \infty$ the horizontal segments $\mathcal{C}_1$ and $\mathcal{C}_2$ essentially reduce to ``two copies'' of the real axis which can be respectively parametrized as $t_1(s) = s$ and $t_2(s) = - s - i \sigma \beta$, with $-\infty < s < \infty$. It is therefore possible to consider the Fourier transform of the propagators $G_{11}, G_{12}, G_{21}$ and $G_{22}$ with respect to the parameter $s$, i.e. 
\begin{equation}\label{Fouriertransform}
G_{i j}(p^0,\mathbf{p}) = \int_{-\infty}^{\infty} d s \, e^{- i p^0 s} \, G_{i j}(t_i(s), {\bf p}) , \qquad\qquad  i, j = 1, 2.
\end{equation}
Notice that these Fourier transforms are well defined because of the exponential term in equation (\ref{propagatorcomponents}). By combining equations (\ref{propagator}), (\ref{propagatorcomponents}) and (\ref{Fouriertransform}) it is easy to derive the following rules for the propagators in the space of momenta \cite{Das:1997gg}:
\begin{subequations} \la{realtimeformalismspropagators}
\ba \la{fey1}
\parbox{20mm}{
\begin{fmfgraph}(65,30) 
\fmfleft{i1}
\fmfright{o1}
\fmf{phantom}{i1,o1}
\fmffreeze
\fmf{vanilla}{i1,o1}
\fmfv{decor.shape=circle, decor.filled=full, decor.size=1.5thick}{i1,o1}
\end{fmfgraph}} \quad = G_{1 1}(p) &=& \frac{i}{p^2 - m^2 + i \epsilon} + 2 \pi n_B (|p^0|) \delta(p^2 - m^2) \\
\parbox{20mm}{
\begin{fmfgraph}(65,30) 
\fmfleft{i1}
\fmfright{o1}
\fmf{phantom}{i1,v,o1}
\fmffreeze
\fmf{vanilla}{i1,v}
\fmf{dashes}{v,o1}
\fmfv{decor.shape=circle, decor.filled=full, decor.size=1.5thick}{i1,o1}
\end{fmfgraph}} \quad = G_{1 2}(p) &=& 2 \pi e^{\sigma \beta p^0} \left[ \theta(- p^0) + n_B (|p^0|)\right] \delta(p^2 - m^2) \\
\parbox{20mm}{
\begin{fmfgraph}(65,30) 
\fmfleft{i1}
\fmfright{o1}
\fmf{phantom}{i1,v,o1}
\fmffreeze
\fmf{vanilla}{v,o1}
\fmf{dashes}{i1,v}
\fmfv{decor.shape=circle, decor.filled=full, decor.size=1.5thick}{i1,o1}
\end{fmfgraph}} \quad = G_{2 1}(p) &=& 2 \pi e^{- \sigma \beta p^0} \left[ \theta( p^0) + n_B (|p^0|)\right] \delta(p^2 - m^2) \\
\parbox{20mm}{
\begin{fmfgraph}(65,30) 
\fmfleft{i1}
\fmfright{o1}
\fmf{phantom}{i1,v,o1}
\fmffreeze
\fmf{dashes}{i1,v,o1}
\fmfv{decor.shape=circle, decor.filled=full, decor.size=1.5thick}{i1,o1}
\end{fmfgraph}} \quad = G_{2 2}(p) &=& - \frac{i}{p^2 - m^2 - i \epsilon} + 2 \pi n_B (|p^0|) \delta(p^2 - m^2) \la{fey4}
\ea
\end{subequations}

After ``splitting'' the propagator $G_\beta$ into the various components defined in equation (\ref{propagatorcomponents}), we can also split the integral along $\mathcal{C}$ appearing in the rule (\ref{vertex}) for the vertex into a sum of integrals along the four segments $\mathcal{C}_i$. However, in the limit $T \to \infty$ only the contributions coming from the segments $\mathcal{C}_1$ and $\mathcal{C}_2$ are relevant, which leads to the following rules for the vertices:
\be
\parbox{20mm}{
\begin{fmfgraph}(45,40)
\fmfleft{i1,i2}
\fmfright{o1,o2}
\fmf{vanilla}{i1,v,o2}
\fmf{vanilla}{i2,v,o1}
\fmfv{decor.shape=circle, decor.filled=full, decor.size=1.5thick}{v}
\end{fmfgraph}
} 
\! \! \! \! = - i g \qquad\qquad\qquad  
\parbox{20mm}{
\begin{fmfgraph}(45,40)
\fmfleft{i1,i2}
\fmfright{o1,o2}
\fmf{dashes}{i1,v,o2}
\fmf{dashes}{i2,v,o1}
\fmfv{decor.shape=circle, decor.filled=full, decor.size=1.5thick}{v}
\end{fmfgraph}}
\! \! \! \! = i g.
\ee
The sign difference between the two vertices is due to the fact that segments $\mathcal{C}_1$ and $\mathcal{C}_2$ have opposite directions, as clearly shown in Figure \ref{figB}. We see therefore that the Feynman rules for real time formalisms are more complicated than the ones for the Matsubara formalism since, for example, four propagators and two vertices are now needed to describe a scalar field with a quartic interaction. This effectively amounts to a doubling of the degrees of freedom necessary to describe the system, and it is a price we need to pay in order to study dynamical properties. 

In conclusion, we should mention that two popular choices for the parameter $\sigma$ appearing in the Feynman rules (\ref{fey1}) -- (\ref{fey4}) for the propagators are $\sigma = 0$ and $\sigma = 1/2$. In the first case, the segments $\mathcal{C}_1$ and $\mathcal{C}_2$ form essentially a closed contour, which explains why the associated formalism is known as \emph{Closed Time Path} (CTP) formalism \cite{Das:1997gg}. The choice $\sigma = 1/2$ reproduces instead the Feynman rules of an operatorial approach to quantum thermal field theory known as \emph{Thermofield Dynamics} (TFD) \cite{Umezawa:1982nv}. Notice that, in this case, the Feynman rules for the propagators become slightly simpler, since it can be easily shown \cite{Das:1997gg} that $G_{12}(p)=G_{21}(p)$.

\section{Hilbert space and Path Integral approaches to Classical mechanics} \la{sec:CPIKvN}

In this section we will briefly review the Hilbert space approach to classical statistical mechanics, which was originally developed by Koopman \cite{Koopman} and von Neumann \cite{vonNeumann1,vonNeumann2} in the 1930s. Then, we shall also present its path integral counterpart \cite{Gozzi:1989bf}. In order to review the main results needed in this paper, we shall restrict ourselves to the simple case of a one-dimensional system, since the extension to systems with more degrees of freedom is straightforward. We refer the reader to the original literature for further details, and in particular to \cite{Carta:2005fq} for an application to classical Yang-Mills theories.

\subsection{Hilbert space approach of Koopman and von Neumann (KvN)} \la{IIa}

The formalism developed by Koopman and von Neumann (KvN) is based on a very simple idea: instead of describing a classical statistical ensemble by means of a probability distribution $\rho(\varphi,t)$ in phase space, where $\varphi\equiv(q,p)$ denotes the phase space variables, one can use a complex   function $\psi(\varphi,t)$ such that the probability distribution is given by the square of its modulus, i.e. 
\begin{equation} \label{eq-1}
\rho(\varphi,t) = |\psi(\varphi,t)|^2.
\end{equation}
Since the probability distribution is normalized, $\psi(\varphi,t)$ must be a  square integrable function and therefore it belongs to a Hilbert space, which will be called \emph{KvN space}. The functions $\psi(\varphi,t)$ will be called instead \emph{KvN waves} in order to distinguish them from the quantum wave functions. Switching to the Dirac notation, $\psi(\varphi,t)$ can be rewritten as $\langle\varphi| \psi, t \rangle$. Note in fact that the eigenstates of the phase-space operators  $|\varphi\rangle = |q, p \rangle$ form a basis of the KvN space. This is due to the fact that, in classical mechanics, the operators $\hat{q}$ and $\hat{p}$ commute, i.e.  
\begin{equation}\label{eq-6}
[ \hat{\varphi}^a, \hat{\varphi}^b] = 0 , 
\end{equation}
and therefore they can be simultaneously diagonalized.

There is however a price to pay in passing from a description of classical ensembles in terms of probability distributions to a description involving states in a Hilbert space. In fact, according to equation (\ref{eq-1}), every distribution $\rho(\varphi,t)$ can be described by a KvN wave of the form:
\begin{equation}
\psi_{\alpha} (\varphi, t) = \sqrt{\rho(\varphi, t)} \, e^{i \alpha(\varphi, t)},
\end{equation}
where $\alpha(\varphi, t)$ is an arbitrary local phase. By varying the local phase we get an infinite set of distinct states in the KvN space which are all associated with the same distribution in phase space, and therefore have the same physical content. Thus, unlike in quantum mechanics, the local phase of a KvN state does not carry any physical information. In fact, it was shown in \cite{DaniloPhD} that the phase and the modulus of a KvN wave satisfy two equations which are completely decoupled and equal to each other. As a consequence, the KvN approach to classical statistical mechanics is somehow redundant, since there is not a  one-to-one correspondence between physical states and waves in the KvN space. As we shall explicitly show in the following sections, this redundancy will allow us to calculate classical thermal averages in many different ways.

In order to better understand this point, let us consider the prescription to calculate statistical averages within the KvN framework. The mean value of a quantity $O(\varphi)$ with respect to the probability distribution $\rho(\varphi,t)$ can be easily expressed in terms of a generic state $|\psi_{\alpha}, t \rangle$ associated with such a distribution in the following way:
\begin{equation} \label{k1}
\langle O(\varphi) \rangle_{\rho} \equiv \langle \psi_{\alpha}, t | O (\hat{\varphi}) | \psi_{\alpha}, t \rangle \equiv \int d \varphi \, O (\varphi) |\langle\varphi|\psi_{\alpha}, t \rangle|^2 = \int d \varphi \, O (\varphi) \rho(\varphi, t),
\end{equation}
where the completeness relation of the eigenstates of $\hat{\varphi}^a$ has been used. More generally, one could also choose to consider a statistical mixture of physically equivalent states $|\psi_{\alpha} \rangle$ described by the ``statistical operator''
\begin{equation} \label{eq-2}
\hat{\rho}(t) \equiv \sum_{\alpha} f_{\alpha} |\psi_{\alpha}, t \rangle \langle \psi_{\alpha}, t |,
\end{equation}
where the real coefficients $f_{\alpha}$ can be chosen arbitrarily. The mean value of $O(\varphi)$ with respect to $\rho(\varphi,t)$ can then be calculated in the following way:
\begin{equation}\label{k2}
\langle O(\varphi) \rangle_{\rho} \equiv \fr{\mbox{Tr} \left[ \hat{\rho}(t) O(\hat{\varphi}) \right]}{\mbox{Tr} \hat{\rho}(t)} =  \fr{\sum_{\alpha} f_{\alpha} \int d \varphi  \,  O (\varphi) |\langle\varphi|\psi_{\alpha}, t \rangle|^2}{\sum_{\alpha} f_{\alpha} \int d \varphi \, | \langle \varphi |\psi_{\alpha}, t \rangle |^2}  = \int d \varphi  \,  O (\varphi) \rho(\varphi, t).
\end{equation}
Thus, every particular choice of the coefficients $f_{\alpha}$ in equation (\ref{eq-2}) results in a different representation of the same statistical distribution $\rho(\varphi, t)$. 

Up to this point we have only considered some general  aspects of the KvN formulation of classical mechanics. Let us now turn to the dynamical aspects and discuss the evolution equation in the KvN space, which is the classical analog of the Schr\"odinger equation. It is well known that every probability distribution $\rho(\varphi,t)$ in phase space evolves in time according to the Liouville equation
\begin{equation}\label{eq-3}
i \partial_t \rho(\varphi, t) = \left( i \partial_q H \partial_p -  i \partial_p H \partial_q \right) \rho(\varphi, t) \equiv \hat{L} \rho(\varphi, t),
\end{equation}
where $H$ is the Hamiltonian of the physical system under consideration and $\hat{L}$ is the generator of time evolution which is called {\it Liouvillian}. The evolution equation for the states $\psi(\varphi, t)$ in the KvN space should be consistent with equations (\ref{eq-1}) and (\ref{eq-3}), and this requirement can be satisfied in a simple way by postulating that $\psi(\varphi, t)$ evolves according to the same equation,~i.e.
\begin{equation}\label{eq-4}
i \partial_t \psi(\varphi, t) = \hat{L}  \psi(\varphi, t).
\end{equation}
Notice that the KvN waves  $\psi(\varphi, t)$ and their square modulus evolve according to the same equation because the Liouvillian is a first-order differential operator. This is not the case in quantum mechanics, where the wave function and its square modulus satisfy different equations because the Hamiltonian is a second-order differential operator.

\subsection{Classical Path Integral (CPI) approach} \la{sec:CPI}

After this  brief introduction to  the Hilbert space approach to classical mechanics, we will now discuss its path integral counterpart and, in particular, we will derive a path integral representation for the kernel 
\begin{equation}\label{eq-12}
K_{f i} \equiv \langle \varphi_f| e^{- i \hat{L} (t_f - t_i)} |\varphi_i \rangle.
\end{equation}
Instead of following the standard slicing method usually adopted in quantum mechanics, we will here proceed in a slightly different and more instructive way. In fact, an implicit expression for the kernel $K_{f i}$ can be derived by resorting to the fact that, in classical mechanics, if we follow the evolution of an eigenstate of $\hat{\varphi}^a$ we get another eigenstate of the same operator.

In order to prove this statement let us first define the derivative operators $\hat{\lambda} = (\hat{\lambda}_q, \hat{\lambda}_p)$ in such a way that
\begin{equation}\label{eq-op5}
\langle \varphi | \hat{\lambda}_a | \psi,t \rangle =  - i \partial_a \psi(\varphi,t).
\end{equation}
These operators satisfy the algebra
\begin{equation}\label{eq-5}
[\hat{\varphi}^a, \hat{\lambda}_b] = i \delta^a_b, \qquad\quad  [\hat{\lambda}_a, \hat{\lambda}_b] = 0 ,
\end{equation}
and can be used in order to rewrite the Liouvillian defined in (\ref{eq-3}) as $\hat{L} = \hat{\lambda}_a \omega^{a b}\partial_b H(\hat{\varphi})$, where $\omega^{a b}$ are the elements of the symplectic matrix \cite{Marsden}. Let us now evolve an eigenstate $| \varphi_i \rangle$ for an infinitesimal amount of time $\varepsilon$ and then consider the action of $\hat{\varphi}^a$ on the resulting state. By using the commutation rules (\ref{eq-6})  and (\ref{eq-5}) we get, up to first-order terms in $\varepsilon$,
\begin{eqnarray}
\hat{\varphi}^a e^{- i \hat{L} \varepsilon} | \varphi_i \rangle &=&  e^{- i \hat{L} \varepsilon} \hat{\varphi}^a | \varphi_i \rangle + \left[ \hat{\varphi}^a, e^{- i \hat{L} \varepsilon}\right]  | \varphi_i \rangle \nonumber \\
 &\approx&  \varphi^a e^{- i \hat{L} \varepsilon}  | \varphi_i \rangle - i \varepsilon \left[ \hat{\varphi}^a, \hat{L} \right] e^{- i \hat{L} \varepsilon} | \varphi_i \rangle \nonumber \\
 &=& \left( \varphi^a_i + \varepsilon \omega^{a b} \partial_b H(\varphi_i)\right) e^{- i \hat{L} \varepsilon} | \varphi_i \rangle \approx \varphi_{cl}^a (\varepsilon; \varphi_i, 0) \, e^{- i \hat{L} \varepsilon} | \varphi_i \rangle \nonumber,
\end{eqnarray}
where $\varphi_{cl}^a (t; \varphi_i, 0)$ is the particular solution of the classical equations of motion $\dot{\varphi}^a = \omega^{a b} \partial_b H(\varphi)$ with initial conditions $\varphi_i$ imposed at $t=0$. Therefore, $e^{- i \hat{L} \varepsilon} | \varphi_i \rangle$ is still an eigenstate of $\hat{\varphi}^a$ and the associated eigenvalue is $\varphi_{cl}^a (\varepsilon; \varphi_i, 0)$. This result is also valid for a finite amount of time and it can be summarized as follows:
\begin{equation}
e^{- i \hat{L} (t_f - t_i)} | \varphi_i \rangle = | \varphi_{cl} (t_f ; \varphi_i, t_i) \rangle.
\end{equation}
As anticipated before, this result can be used to find an implicit expression for the  kernel $K_{f i}$ in terms of the solution to the equations of motion. In fact, by exploiting the orthonormality of the eigenstates of $\hat{\varphi}^a$ we get:
\begin{equation}\label{eq-7}
K_{f i} = \langle \varphi_f| e^{- i \hat{L} (t_f - t_i)} |\varphi_i \rangle = \langle \varphi_f | \varphi_{cl} (t_f ; \varphi_i, t_i) \rangle = \delta \left(  \varphi_f -  \varphi_{cl} (t_f ; \varphi_i, t_i)\right).
\end{equation}
This expression admits a simple physical interpretation, since it basically states that the classical probability amplitude of going from $\varphi_i$ to $\varphi_f$ in a time interval $t_f - t_i$ is different than zero only if the classical trajectory coming out from $\varphi_i$ at time $t_i$ passes through $\varphi_f$ at time $t_f$. Of course, equation (\ref{eq-7}) should be regarded as an implicit result, since in most cases we are not able to explicitly solve the equations of motion. It is however a good starting point for deriving a path integral representation of $K_{f i}$. 

To this end, let us first rewrite the kernel as a functional integral of a functional Dirac delta involving the whole classical trajectory.  This can be done by slicing the time interval $t_f - t_i$ in $N+1$ subintervals of lenght $\varepsilon$ and taking the limit $N \rightarrow \infty$:
\begin{eqnarray}\label{eq-8}
\delta (\varphi_f - \varphi_{cl}(t_f ; \varphi_i, t_i)) &=& \prod_{n=1}^N \int d \varphi_n \prod_{n=0}^N \delta(\varphi_{n+1} - \varphi_{cl}(t_{n+1} ; \varphi_n, t_n)) \nonumber \\
&\mbox{$\stackrel{N \rightarrow \infty}{\longrightarrow}$}& \int \mathcal{D}'' \varphi \, \, \delta (\varphi(t) - \varphi_{cl}(t ; \varphi_i, t_i)). \label{15.2}
\end{eqnarray}
In deriving this result we have defined $t_n \equiv t_i + n \varepsilon$ and $\varphi_n \equiv \varphi(t_n)$. Furthermore, $\mathcal{D}'' \varphi$ indicates that the functional integration is over all the paths $\varphi(t)$ with fixed end points $\varphi_i$ and $\varphi_f$. The functional Dirac delta in equation (\ref{eq-8}) can be expressed in terms of another delta function involving the Hamilton equations of motion:
\begin{equation}\label{eq-11}
 \delta (\varphi^a - \varphi_{cl}^a) = \delta (\dot{\varphi}^a - \omega^{a b} \partial_b H) \, \mbox{det} (\delta^a_b \partial_t - \omega^{a c} \partial_c \partial_b H).
 \end{equation}
Both elements on the right-hand side of the above equation can be expressed in terms of functional integrals. For the first one we can use the functional Fourier representation
\begin{equation}\label{eq-9}
\delta (\dot{\varphi}^a - \omega^{a b} \partial_b H) = \int \mathcal{D} \lambda \exp \left\{ i \int_{t_i}^{t_f} dt \, \lambda_a \left( \dot{\varphi}^a - \omega^{a b} \partial_b H \right) \right\},
\end{equation}
while the second one, using four Grassmann variables $c^a = (c^q,c^p)$ and $\bar{c}_a = (\bar{c}_q, \bar{c}_p)$, can be expressed as follows:
\begin{equation}\label{eq-10}
\mbox{det} (\delta^a_b \partial_t - \omega^{a c} \partial_c \partial_b H) = \int \mathcal{D} \bar{c}  \, \mathcal{D} c \, \exp \left\{ - \int_{t_i}^{t_f} dt \, \bar{c}_a (\delta^a_b \partial_t - \omega^{a c} \partial_c \partial_b H) c^b \right\}.
\end{equation}
By combining equations (\ref{eq-7}) through (\ref{eq-10}) we obtain the following path integral representation of $K_{f i}$:
\begin{equation} \label{eq-13}
K_{f i} =  \int \mathcal{D}'' \varphi \, \mathcal{D} \lambda \, \mathcal{D} \bar{c}  \, \mathcal{D} c \, \exp \left\{ i \int_{t_i}^{t_f} dt \left( \lambda_a \dot{\varphi}^a + i \bar{c}_a \dot{c}^a - \mathcal{H} \right) \right\},
\end{equation}
where we have defined 
\begin{equation} \label{eq-31}
\mathcal{H} \equiv \lambda_a \omega^{a b} \partial_b H + i \bar{c}_a \omega^{a b} \partial_b \partial_d H c^d.
\end{equation}
The first term in $\mathcal{H}$ is just the Liouvillian $L$, while the second term is responsible for the time evolution of the Grassmann variables $c^a$ and $\bar{c}_a$ \cite{Gozzi:1989bf}. 

It was shown in \cite{Gozzi:1989bf} that the variables $c^a$ turn out to have an interesting physical interpretation in terms of Jacobi fields, which are the small variations among nearby classical trajectories. As a consequence, their correlation functions are related to the Lyaupunov exponents of the system described by the Hamiltonian $H$ \cite{Kurchan03}. Moreover, these variables can also be interpreted as a basis of differential 1-forms $d\varphi^a$ \cite{Gozzi:1989bf,Gozzi:1999uq}. Indeed, notice that, due to the anticommuting nature of these variables, the antisymmetry property of the wedge product between 1-forms is automatically recovered. Similarly, $\mathcal{H}$ itself corresponds to the Lie derivative of the Hamiltonian flow~\cite{Marsden} associated with the Hamiltonian~$H$. Therefore, since the Grassmann variables $c^a$ carry some interesting mathematical and physical information, it is important to consider the extended Hilbert space \cite{Deotto:2002hy} associated with the path-integral~(\ref{eq-13}). 

From the kinetic term in (\ref{eq-13}), it is easy to see that the operators $\hat{c}^a$ and $\hat{\bar{c}}_a$ acting on this extended space must obey the following algebra \cite{Gozzi:1989bf}:
\begin{equation} \label{eq-30}
 \left\{ \hat{c}^a, \hat{\bar{c}}_b \right\} = \delta^a_b, \qquad \left\{ \hat{c}^a, \hat{c}^b \right\} = \left\{ \hat{\bar{c}}_a, \hat{\bar{c}}_b\right\} = 0 \ ,
\end{equation}
where the brackets above denote anti-commutators.
Furthermore, $\hat{c}^a$ and $\hat{\bar{c}}_a$ commute with the operators $\hat{\varphi}^a$ and $\hat{\lambda}_a$. In this enlarged Hilbert space, time evolution is generated by the operator $\hat{\mathcal{H}}$ defined in equation (\ref{eq-31}), and every state obeys the equation of motion~\cite{Gozzi:1989bf}
\begin{equation}
i \partial_t  | \psi, t \rangle = \hat{\mathcal{H}} \, | \psi, t \rangle,
\end{equation}
which is a generalization of equation (\ref{eq-4}). The evolution generated by this equation may or may not be unitary, depending on how the scalar product is defined. In what follows, we will adopt the \emph{symplectic scalar product} \cite{Deotto:2002hy} such that $(\hat{c}^a)^\dag = i \omega^{a b} \hat{\bar{c}}_b$. With this definition, it is easy to see that the operator $\hat{\mathcal{H}}$ is Hermitian, and therefore that time evolution is unitary. A brief review of the symplectic scalar product can be found in the Appendix \ref{app:symplectic}, while we refer the reader to \cite{Deotto:2002hy} for more details concerning alternative implementations of the scalar product.

\subsection{Universal supersymmetry and superfields} \la{SUSY}

One striking feature of the approach to classical mechanics we outlined above is the large number of auxiliary variables. In fact, while the usual Hamiltonian formulation requires $2n$ variables $\varphi^a$ to describe a physical system with $n$ degrees of freedom, the path integral approach (or, equivalently, the formulation based on the extended Hilbert space) makes use of the $8n$ variables $\varphi^a, \lambda_a, c^a$ and $\bar{c}_a$. However, the redundancy of this description is accompanied by a high degree of internal symmetry, as already noticed in \cite{Gozzi:1989bf,Gozzi:1989xz}. In what follows, we will mostly be interested in the symmetry transformations associated with the following charges \cite{Gozzi:1989xz}:
\begin{eqnarray}
&& \hat{Q}_H = \hat{c}^a \left( i \hat{\lambda}_a + \beta \partial_a H(\hat{\vphi})/ 2\right) \\
&& \hat{\bar{Q}}_H = \hat{\bar{c}}_a \omega^{a b} \left( i \hat{\lambda}_b - \beta \partial_b H (\hat{\vphi})/ 2 \right),
\end{eqnarray}
where $\beta$ is an arbitrary parameter. By using the algebra given by the equations (\ref{eq-6}), (\ref{eq-5}) and (\ref{eq-30}), it is easy to check that these charges commute with the generator of time evolution $\hat{\mathcal{H}}$ and therefore that they are conserved. Furthermore, by using the Hermiticity conditions $(\hat{c}^a)^\dag = i \omega^{a b} \hat{\bar{c}}_b$ one can show that $\hat{Q}_{H}^{\dag} = i \, \hat{\bar{Q}}_{H}$, and thus that the operator
\begin{equation}
\hat{U}  (\epsilon, \bar{\epsilon}) = \exp \left(- \epsilon \, \hat{Q}_{H} \, - \, \hat{\bar{Q}}_{H} \bar{\epsilon} \right) 
\end{equation}
generates a unitary transformation (provided $ \epsilon^* = i \bar{\epsilon}$). In order to better understand the nature of this symmetry, we can consider the algebra obeyed by the charges $\hat{Q}_H$ and $\hat{\bar{Q}}_H$, which reads
\begin{eqnarray}\label{eq-t.3}
\left\{ \hat{Q}_H,  \hat{\bar{Q}}_H \right\} = - i \beta \, \hat{\mathcal{H}}.
\end{eqnarray}
From this equation, we can see that $\hat{Q}_H$ and $\hat{\bar{Q}}_H$ are the generators of a non-relativistic N=2 supersymmetry (SUSY). Notice that this symmetry has a universal character, in the sense that it does not depend on the nature of the physical system, i.e. on the form of the Hamiltonian $H$. 

This universal supersymmetry allows a simplification of the formalism  by combining the variables $\varphi^a$, $\lambda_a$, $c^a$ and $\bar{c}_a$ into a single object which bears a close resemblance to the superfields one can introduce in ordinary supersymmetric field theories. As it is well known, the introduction of a superfield requires us to enlarge the base space. In our case, this means that, besides the ordinary time $t$, two ``Grassmannian'' partners of time $\theta$ and $\bar{\theta}$ are needed. The triplet $\tau \equiv (t,\theta,\bar{\theta})$ is called {\it supertime} and plays the same role as the superspace in supersymmetric field theories. A superfield can then be introduced as follows  \cite{Abrikosov:2004cf}:
\be \la{superfield}
\displaystyle \Phi^a(t,\theta,\bar{\theta})\equiv \varphi^a(t)+\theta c^a(t)+\bar{\theta}\omega^{ab}\bar{c}_b(t)+i\bar{\theta}\theta \omega^{ab}\lambda_b(t).
\ee
With our convention for the scalar product\footnote{We are also \emph{defining} the complex conjugate of a product of Grassmann numbers as $ (\theta_1 \theta_2)^* \equiv \theta_2^* \theta_1^*$.}, it immediately follows that $\Phi^a$ is Hermitian provided $\theta^* = i \bar{\theta}$.

Many of the results presented above can be expressed much more elegantly in terms of superfields. For example, the algebra given in equations (\ref{eq-6}), (\ref{eq-5}) and (\ref{eq-30}) is fully encoded in the much more compact equation \cite{Abrikosov:2004cf}
\begin{equation}
\left[ \hat{\Phi}^a(t,\theta,\bar{\theta}) , \hat{\Phi}^b(t,\theta',\bar{\theta}') \right] = i \omega^{ab} \delta(\theta - \theta') \delta(\bar{\theta} - \bar{\theta}'),
\end{equation}
where we remind the reader that, for a Grassmann variable $\theta$, the Dirac delta is simply given by $\delta(\theta) = \theta$. A great simplification occurs also at the level of the path integral. In fact, let us consider the Hamiltonian $H(\varphi)$ and replace the phase space variables $\varphi$ with the superfields $\Phi$. By expanding $H(\Phi)$ in powers of $\theta$ and $\bar{\theta}$, we obtain
\begin{equation} \label{multiplet}
\displaystyle H(\Phi)=H(\varphi)+\theta N+\bar{N}\bar{\theta}-i\bar{\theta}\theta {\mathcal{H}}, 
\end{equation}
where the two components $N$ and $\bar{N}$ of the supermultiplet above are two functions of $\varphi^a$, $c^a$, $\bar{c}_a$ and $\lambda_a$ whose explicit form is not relevant to our analysis. On the other hand, the last component of the multiplet is the generator of time evolution defined in equation (\ref{eq-31}), which can be extracted from $H(\Phi)$ by an integration over the Grassmann partners of time $\theta$ and $\bar{\theta}$:
\begin{equation}
{\mathcal{H}}=i\int d\theta d\bar{\theta}\, H(\Phi).
\end{equation}
Similarly, the kinetic term in equation (\ref{eq-13}) can be rewritten by introducing an integral over $\theta$ and $\bar{\theta}$ as follows:
\begin{equation}
\lambda_a\dot{\varphi}^a+i\bar{c}_a\dot{c}^a= i \int d\theta d\bar{\theta}\, \Phi^p\dot{\Phi}^q+\frac{d}{dt}(\lambda_pp+i\bar{c}_pc^p).
\end{equation}
Because of the surface terms appearing on the LHS of this equation, the path integral (\ref{eq-13}) cannot be expressed solely in terms of superfields. However, these surface terms disappear if, instead of considering $K_{fi}$ (which is essentially the transition amplitude between two states $|\varphi_i, c_i \rangle$  and $| \varphi_f, c_f \rangle$ integrated over the initial and final values of $c^a$), we calculate the transition amplitude between two eigenstates of the superfield $\hat{\Phi}^q=\hat{q}+\theta \hat{c}^q+\bar{\theta}\hat{\bar{c}}_p+i\bar{\theta}\theta \hat{\lambda}_p$. Notice in fact that all the components of $\hat{\Phi}^q$ commute among each other and therefore they can be diagonalized simultaneously (see \cite{Abrikosov:2004cf} or Appendix A for more details). In conclusion, we have \cite{Abrikosov:2004cf}
\begin{equation}
\langle \Phi^q_f, t_f | \Phi^q_i,t_i\rangle = \int {\mathcal D}^{\prime\prime}\Phi^q {\mathcal D}\Phi^p \, \exp \left\{ i\int_{t_i}^{t_f} d\tau \, \left[\Phi^p\dot{\Phi}^q-H(\Phi)\right] \right\}, \label{classical}
\end{equation}
where we have denoted with $|\Phi^q\rangle$ the eigenstates of $\hat{\Phi}^q$, and we have defined the differential of the supertime $\tau = (t,\theta, \bar{\theta})$ as $d \tau \equiv i \, dt \, d\theta d\bar{\theta}$. Notice that, according to our conventions, $d \tau$ is real. Furthermore, we used  the notation $\mathcal{D}'' \Phi^q$ to indicate that we are integrating over all the paths $\Phi^q(t)$ with fixed end points $\Phi^q_i$ and $\Phi^q_f$.

The formal similarity between the result (\ref{classical}) and the standard path integral expression for a transition amplitude in quantum mechanics is striking:
\begin{equation}
\langle q_f, t_f | q_i,t_i\rangle = \int {\mathcal D}^{\prime\prime}q {\mathcal D}p \, \exp \left\{ i \int_{t_i}^{t_f} dt \, \left[p\dot{q}-H(\varphi)\right] \right\}. \label{quantum}
\end{equation}
It is easy to realize \cite{Abrikosov:2004cf} that the ``dequantization rules'' which let the quantum transition amplitude (\ref{quantum}) turn into the classical one (\ref{classical}) are the following:
\begin{itemize}
\item[1)] Replace the phase space variables $(q,p)$ with the associated superfields $(\Phi^q,\Phi^p)$;
\item[2)] Extend the time integration to a supertime integration
\begin{displaymath}
\displaystyle \int dt \, \longrightarrow \,  \int d\tau.
\end{displaymath}
\end{itemize}
As we will see, in order to establish a connection between classical and quantum field theory in thermal equilibrium, we will need to supplement these two rules with a third one, which will be derived in section \ref{classical matsubara path integral}.

\section{Classical CTP Formalism} \la{sec:classicalCTP}

We will now start applying the classical formalism reviewed in the previous section to the study of classical thermal field theory (CTFT). In particular, we will first review the path integral formulation of CTFT developed in \cite{Jeon:2004dh}. As we will see, this approach turns out to be related to the high temperature limit of the CTP formulation of quantum thermal field theory. 

In this section as well as in the following ones, we will continue to illustrate the general formalism by considering the concrete example of a scalar field $\phi$ described by the Hamiltonian
\begin{equation} \label{eq-40}
H = \int d^3 x \left\{ \frac{\pi^2}{2} + \frac{(\nabla \phi)^2}{2} + \frac{m^2 \phi^2}{2}  + \fr{g \, \phi^4}{4!} \right\} .
\end{equation}
Of course, our results can be extended to more general systems in a straightforward way. In analogy with the previous section, we will indicate the collection of the phase space variables as $\varphi = (\phi, \pi)$. 

The starting point of the approach developed in \cite{Jeon:2004dh} is the following expression for the $n$-point thermal correlation functions \cite{Aarts:1996qi, Aarts:1997kp}:
\begin{equation} \label{p.9}
\langle \phi (x_1) \, ... \, \phi (x_n) \rangle_{\beta} \equiv Z^{-1}_{\beta} \int [d \varphi_i] \, \phi_{cl} (x_1; \varphi_i, t_i) \, ... \, \phi_{cl} (x_n ; \varphi_i, t_i) \, e^{- \beta H(\varphi_i)}.
\end{equation}
In this formula, $\phi_{cl} (x ; \varphi_i, t_i)$ is the solution of the classical equation of motion associated with the Hamiltonian (\ref{eq-40}),  $[d \varphi_i]$ denotes a path integral over all the possible field configurations at a fixed time $t_i$, and $Z_{\beta}$ is the classical canonical partition function defined as
\begin{equation}
Z_{\beta} = \int [d \varphi] \, e^{- \beta H(\varphi)}.
\end{equation}
Equation (\ref{p.9}) is just the standard definition of thermal average, and it basically states that $\langle \phi (x_1) \, ... \, \phi (x_n) \rangle_{\beta}$ is given by the product of the classical solutions evaluated at points $x_1, ..., x_n$ with initial conditions $\varphi_i$ averaged over the Boltzmann distribution.

\subsection{Generating functional}

Equation (\ref{p.9}) can be easily re-expressed in terms of an expectation value in the KvN space. In fact, the solution of the classical equations of motion can be written as:
\begin{equation}\label{p.1}
\phi_{cl} (x; \varphi_i, t_i) = \int [d \varphi_f] \, \langle \varphi_f, t_f | \hat{\phi}(x) | \varphi_i, t_i \rangle,
\end{equation}
where $x = (t, {\bf x})$ and $ t_i \leq t \leq t_f$. This result can be easily proven in a few steps. In fact, dropping momentarily the dependence on the space variable ${\bf x}$, we have:
\begin{eqnarray}
&& \int [d \varphi_f] \, \langle \varphi_f, t_f | \hat{\phi}(t) | \varphi_i, t_i \rangle = \int [d \varphi_f] [d \varphi] \, \langle \varphi_f, t_f | \varphi, t \rangle \, \phi \, \langle \varphi, t |  \varphi_i, t_i \rangle \nonumber \\
&& \qquad \qquad \qquad = \int [d \varphi_f] [d \varphi] \, \delta \left( \varphi_f - \varphi_{cl}(t_f ; \varphi, t) \right) \, \phi \, \delta \left( \varphi - \varphi_{cl}(t ; \varphi_i, t_i) \right) \nonumber \\
&&  \qquad \qquad \qquad = \int  [d \varphi_f] \, \phi_{cl} (t; \varphi_i, t_i) \, \delta \left( \varphi_f - \varphi_{cl}(t_f ; \varphi_{cl}(t ; \varphi_i, t_i), t) \right) \nonumber \\
&&  \qquad \qquad \qquad = \phi_{cl} (t; \varphi_i, t_i), \nonumber
\end{eqnarray}
where in the first line we have used the completeness of the eigenstates $|\varphi, t \rangle$ and in passing to the second line we have rewritten the transition amplitudes by using equation (\ref{eq-7}). Similarly, it is easy to show that the thermal correlation functions introduced in (\ref{p.9}) can be rewritten as
\begin{equation} \label{G-ctp-operatorial}
\langle \phi (x_1) \, ... \, \phi (x_n) \rangle_{\beta} = Z^{-1}_{\beta} \int [d \varphi_f] [d \varphi_i] \, \langle \varphi_f, t_f | T \left[ \hat{\phi}(x_1) \, ... \,   \hat{\phi}(x_n) \right] | \varphi_i, t_i \rangle \, e^{- \beta H(\varphi_i)},
\end{equation}
where $t_i \leq t_k \leq t_f$ for any $k= 1, ..., n$.
Equation (\ref{G-ctp-operatorial}) establishes a connection between the standard approach to the canonical ensemble encoded in equation (\ref{p.9}) on one hand, and the KvN operatorial approach on the other hand. By resorting to a slicing procedure familiar from ordinary quantum field theory and by using the result (\ref{eq-13}), equation (\ref{G-ctp-operatorial}) can also be expressed as a ratio of path integrals:
\be \la{starting point}
\langle \phi (x_1) \, ... \, \phi (x_n) \rangle_{\beta} = \frac{ \int \mathcal{D} \varphi \, \mathcal{D} \lambda \, \mathcal{D} \bar{c} \, \mathcal{D} c\,\, \phi (x_1) \, ... \, \phi (x_n) \, \exp \left\{ i \, \mathcal{S}  - \beta H(\vphi(t_i)) \right\}}{\int \mathcal{D} \varphi \, \mathcal{D} \lambda\, \mathcal{D} \bar{c} \, \mathcal{D} c\, \exp \left\{ i \, \mathcal{S} - \beta H(\vphi(t_i)) \right\}} 
\ee
where, according to equations (\ref{eq-31}) and (\ref{eq-40}), we have that  $\mathcal{S} = \mathcal{S}_1 + \mathcal{S}_2 + \mathcal{S}_3$ with
\ba
\mathcal{S}_1 &=& \int_{t_i}^{t_f} dt \int d^3 x \l[ \lambda_\phi \dot{\phi} + \lambda_\pi \dot{\pi} - \lambda_\phi \pi +\lambda_\pi \l(- \nabla^2 + m^2 \r) \phi \r] \la{S1}\\
\mathcal{S}_2 &=& \int_{t_i}^{t_f} dt \int d^3 x \l[  i \bar{c}_\phi \dot{c}^\phi +  i \bar{c}_\pi \dot{c}^\pi - i \bar{c}_\phi c^\pi + i \bar{c}_\pi \l(- \nabla^2 + m^2 \r) c^\phi \r] \la{S2}\\
\mathcal{S}_3 &=& \int_{t_i}^{t_f} dt \int d^3 x \l[ \frac{g}{3!} \lambda_\pi \, \phi^3 +  \frac{i g}{2} \, \bar{c}_\pi c^\phi  \phi^2\r] . \la{S3}
\ea
Notice that we have decomposed $\mathcal{S}$ into a part  $\mathcal{S}_1$,which is quadratic in ($\vphi^a, \lambda_a$),  a part $\mathcal{S}_2$ which is quadratic in the Grassmann variables ($c^a, \bar{c}_a$), and a part $\mathcal{S}_3$ which is quartic in the fields  and introduces a coupling among the various fields. In particular, the fact that $\phi$ is coupled to the auxiliary variables $\lambda_\pi, c^\phi$ and $\bar{c}_\pi$, suggests that a perturbative calculation of thermal correlation functions of the field $\phi$ based on the expression (\ref{starting point}) requires the introduction of a generating functional $Z$ which depends on four currents, namely
\be \la{Jeon's generating functional}
Z \equiv  \frac{\displaystyle \int \mathcal{D} \varphi \, \mathcal{D} \lambda \, \mathcal{D} \bar{c}\, \mathcal{D} c \,\exp \left\{ i \, \mathcal{S} - \beta H(\vphi(t_i))  + i \int \l( \Lambda^\pi \lambda_\pi + J_\phi \phi - i \bar{\eta}_\phi c^\phi - i \bar{c}_\pi \eta^\pi\r) \right\}}{\displaystyle \int \mathcal{D} \varphi \, \mathcal{D} \lambda\, \mathcal{D} \bar{c}  \, \mathcal{D} c\, \exp \left\{ i \, \mathcal{S} - \beta H(\vphi(t_i)) \right\}} .
\ee
Clearly, the thermal average of a product of $n$ fields $\phi$ can be calculated by taking $n$ derivatives of $Z$ with respect to $J_\phi$ and eventually setting the external currents equal to zero:
\begin{equation} \la{derivatives wrt Jphi}
\langle \phi (x_1) \, ... \, \phi (x_n) \rangle_{\beta} = \left( \frac{1}{i} \right)^n \left. \frac{\delta^{n} Z}{ \delta J_{\phi}(x_1) \, ... \, \delta J_{\phi} (x_n)} \right|_{J_{\phi}, \Lambda^\pi, \bar{\eta}_\phi, \eta^\pi = 0} .
\end{equation}

\subsection{Feynman rules} \label{ss:1}

We can now use (\ref{Jeon's generating functional}) to derive a set of Feynman rules which will allow us to calculate thermal correlation functions in a systematic way. The propagators can be easily calculated by considering the case in which the coupling $g$ in equation (\ref{eq-40}) vanishes and, therefore, the quartic terms included in $\mathcal{S}_3$ are absent. In this case, the Grassmann variables $c^a,\bar{c}_a$ decouple from $\vphi^a$ and $\lambda_a$, and the generating functional (\ref{Jeon's generating functional}) reduces to the product of two terms, $Z= Z_1 [J_\phi, \Lambda^\pi] \, Z_2 [\bar{\eta}_\phi, \eta^\pi]$, with
\ba
Z_1[J_\phi, \Lambda^\pi] &=& \frac{\displaystyle\int \mathcal{D} \varphi \, \mathcal{D} \lambda \, \exp \left\{  i \, \mathcal{S}_1 - \beta H(\vphi(t_i))  + i \int \l( \Lambda^\pi \lambda_\pi + J_\phi \phi \r) \right\}}{\displaystyle \int \mathcal{D} \varphi \, \mathcal{D} \lambda \,\exp \left\{ i \, \mathcal{S}_1 - \beta H(\vphi(t_i)) \right\}} \\
Z_2 [\bar{\eta}_\phi, \eta^\pi] &=& \frac{\displaystyle\int \mathcal{D} c\, \mathcal{D} \bar{c} \, \exp \left\{  i \, \mathcal{S}_2  + \int \l(  \bar{\eta}_\phi c^\phi + \bar{c}_\pi \eta^\pi \r) \right\}}{\displaystyle \int \mathcal{D} c \, \mathcal{D} \bar{c}\,\exp \left\{ i \, \mathcal{S}_2 \right\}} .
\ea
Both these quantities involve only Gaussian integrals and therefore can be calculated exactly. In particular, $Z_1$ can be easily calculated by using the result  
\ba 
&&\int d\vphi_f \int \mathcal{D}'' \varphi \, \mathcal{D} \lambda \, \exp \l\{ i \mathcal{S}_1 + i \int \l( \Lambda^\pi \lambda_\pi + J_\phi \phi \r) \r\} = \label{penco mauro result}\\
&& \qquad \qquad \qquad = \mathcal{N} \exp \left\{ i \int_{t_i}^{t_f} d^4 x \, J_{\phi} (x) \phi_0 (x) + i  \int_{t_i}^{t_f} d^4 x \, d^4 x^{\prime} J_{\phi} (x) G_R (x - x^{\prime}) \Lambda^{\pi} (x^{\prime}) \right\},  \nonumber 
\ea
where $\phi_0(x)$ denotes the solution of the \emph{free} equations of motion with no currents and with initial conditions $\varphi_i$ at time $t_i$. More explicitly, we have
$$
\phi_0 (x ; \varphi_i, t_i) = \int \frac{d^3 p}{(2 \pi)^3} \left( \phi_i ({\bf p}) \cos [ E_{{\bf p}}(t - t_i)] + \frac{\pi_i({\bf p})}{E_{{\bf p}}} \sin [ E_{{\bf p}}(t - t_i)]\right) e^{i {\bf p \cdot x}},
$$ 
where again $E_{{\bf p}} = \sqrt{{\bf p}^2 + m^2}$, while  $G_R(x)$ is the retarded Green function, which admits the following Fourier representation \cite{Das:1997gg}:
\begin{equation}
G_R (x) = \int \frac{d^4 p}{(2 \pi)^4} \, \frac{e^{- i p \cdot x}}{p^2 - m^2 +i \epsilon p^0}.
\end{equation}
For a derivation of equation (\ref{penco mauro result}), we refer the reader to \cite{Penco:2006wn}, where an analogous result was derived in the simple case of a single harmonic oscillator. At this point, all we need to do in order to calculate $Z_1$ is to integrate (\ref{penco mauro result}) over the initial conditions $\vphi_i$ weighted by the Boltzmann distribution and impose the normalization condition $Z_1=1$ when the external currents vanish. This is just a Gaussian integral which can be calculated exactly, and the final result becomes particularly simple in the limit where $t_i \rightarrow - \infty$ and $t_f \rightarrow  \infty$. In this case, we simply have
\be \la{Z1}
Z_1 = \exp \left\{ - \frac{1}{2} \int d^4x \, d^4x^{\prime} \left[ J_{\phi}(x) \Delta_{\beta} (x - x^{\prime}) J_{\phi}(x^{\prime})  - 2 \, i J_{\phi} (x) G_R (x - x^{\prime}) \Lambda^{\pi} (x^{\prime}) \right] \right\}
\ee
where 
\begin{equation} \la{delta beta}
\Delta_{\beta} (x) \equiv  \int \frac{d^4 p}{(2 \pi)^4} \, \frac{2 \pi}{ \beta | p^0|} \,  \delta (p^2 - m^2) \, e^{- i p \cdot x}.
\end{equation}

The calculation of $Z_2$ is somehow  subtle, and the details can be found in the Appendix~\ref{app: ghost propagator}, but the final result is very simple:
\begin{equation} \la{Z2}
Z_2 = \exp \left\{ - \int d^4 x \, d^4 x^{\prime} \, \bar{\eta}_\phi (x) G_R(x - x^{\prime}) \eta^\pi (x^{\prime}) \right\}.
\end{equation}
From equations (\ref{Z1}) and (\ref{Z2}) we can immediately read off the Feynman rules for the propagators in the momentum space:
\be \label{eq-60}
\begin{array}{ll}
G_{ \phi \phi} = \, \parbox{20mm}{
\begin{fmfgraph}(65,30) 
\fmfleft{i1}
\fmfright{o1}
\fmf{phantom}{i1,o1}
\fmffreeze
\fmf{vanilla}{i1,o1}
\fmfv{decor.shape=circle, decor.filled=full, decor.size=1.5thick}{i1,o1}
\end{fmfgraph}} \quad = \Delta_{\beta} (p) &
\qquad 
G_{ \phi \lambda_\pi} = \, \parbox{20mm}{
\begin{fmfgraph}(65,30) 
\fmfleft{i1}
\fmfright{o1}
\fmf{phantom}{i1,v,o1}
\fmffreeze
\fmf{vanilla}{i1,v}
\fmf{dashes}{v,o1}
\fmfv{decor.shape=circle, decor.filled=full, decor.size=1.5thick}{i1,o1}
\end{fmfgraph}} \quad = - i G_R (p) \\
G_{ \lambda_\pi \phi} = \, \parbox{20mm}{
\begin{fmfgraph}(65,30) 
\fmfleft{i1}
\fmfright{o1}
\fmf{phantom}{i1,v,o1}
\fmffreeze
\fmf{vanilla}{v,o1}
\fmf{dashes}{i1,v}
\fmfv{decor.shape=circle, decor.filled=full, decor.size=1.5thick}{i1,o1}
\end{fmfgraph}} \quad = - i G_A (p) & \qquad
G_{ \lambda_\pi \lambda_\pi} = \, \parbox{20mm}{
\begin{fmfgraph}(65,30) 
\fmfleft{i1}
\fmfright{o1}
\fmf{dashes}{i1,o1}
\fmfv{decor.shape=circle, decor.filled=full, decor.size=1.5thick}{i1,o1}
\end{fmfgraph}} \quad = 0 \\
G_{ c^\phi \bar{c}_\pi} = \, \parbox{20mm}{
\begin{fmfgraph}(65,30) 
\fmfleft{i1}
\fmfright{o1}
\fmf{ghost}{i1,o1}
\fmfv{decor.shape=circle, decor.filled=full, decor.size=1.5thick}{i1,o1}
\end{fmfgraph}} \quad = - G_R (p) & \qquad 
G_{  \bar{c}_\pi c^\phi} = \, \parbox{20mm}{
\begin{fmfgraph}(65,30) 
\fmfleft{i1}
\fmfright{o1}
\fmf{ghost}{o1,i1}
\fmfv{decor.shape=circle, decor.filled=full, decor.size=1.5thick}{i1,o1}
\end{fmfgraph}} \quad =  G_A (p),
\end{array}
\ee
where we introduced the advanced Green function which, in momentum space, satisfies the relation $G_A(p) \equiv G_R(-p)$.

Let us now consider the quartic interactions in (\ref{eq-40}), which in principle will affect equation (\ref{Jeon's generating functional}) not only through $\mathcal{S}_3$ (see equation (\ref{S3})), but also through the Boltzmann weight $- \beta H(\vphi(t_i))$. However, the latter dependence becomes negligible in the limit $t_i \to - \infty$ provided \cite{Aarts:1997kp} we turn off the potential adiabatically by modifying the coupling $g$ as follows:
\begin{equation}
g \rightarrow g \, e^{- \epsilon |t|},
\end{equation}
with $\epsilon$ a positive infinitesimal parameter. Then, the only contribution will come from $\mathcal{S}_3$, and the Feynman rules for the vertices in momentum space can be simply deduced from the expression (\ref{S3}):
\begin{equation} \la{classical CTP vertices}
\parbox{20mm}{
\begin{fmfgraph}(45,40)
\fmfleft{i1,i2}
\fmfright{o1,o2}
\fmf{vanilla}{i1,v,o1}
\fmf{vanilla}{i2,v}
\fmf{dashes}{v,o2}
\fmfv{decor.shape=circle, decor.filled=full, decor.size=1.5thick}{v}
\end{fmfgraph}} \! \! \! \! = i g \qquad \qquad
\parbox{20mm}{
\begin{fmfgraph}(45,40)
\fmfleft{i1,i2}
\fmfright{o1,o2}
\fmf{ghost}{i1,v,o1}
\fmf{vanilla}{i2,v,o2}
\fmfv{decor.shape=circle, decor.filled=full, decor.size=1.5thick}{v}
\end{fmfgraph}} \! \! \! \! = - g.
\end{equation}

It is instructive to apply the Feyman rules derived above to calculate the one loop corrections to the 2-point function $\langle \phi(x_1) \phi (x_2) \rangle_{\beta}$ in momentum space, which are represented by the following diagrams:
\be \la{one loop diagrams}
\parbox[b]{20mm}{\includegraphics[scale=0.4]{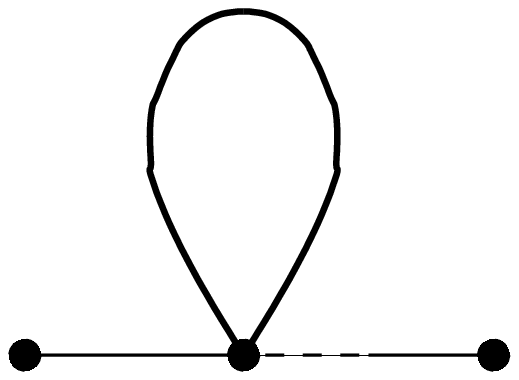}} \quad + \quad  \parbox[b]{20mm}{\includegraphics[scale=0.4]{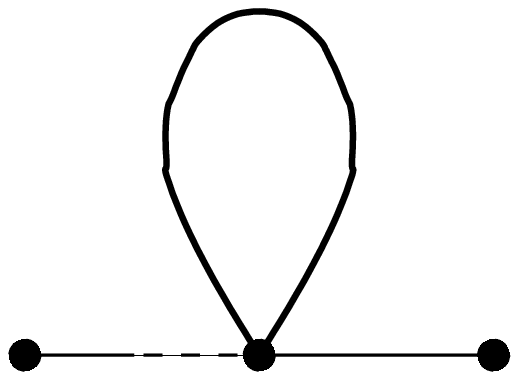}} \quad + \quad
\parbox[b]{20mm}{\includegraphics[scale=0.4]{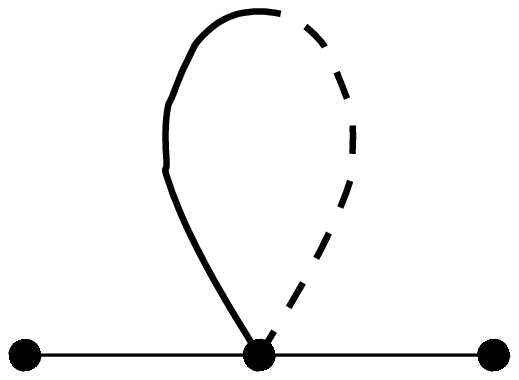}} \quad + \quad
\parbox[b]{20mm}{\includegraphics[scale=0.4]{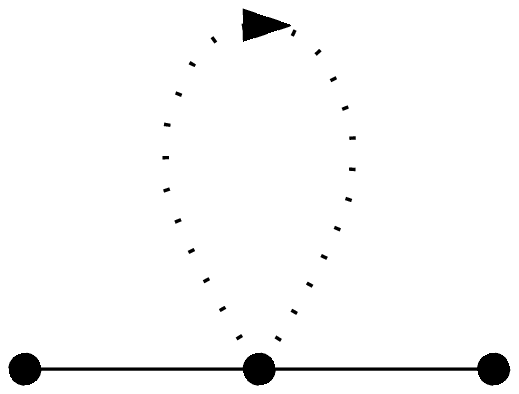}} \: .
\ee
The contribution of the first two graphs is essentially determined by the following loop:
\be \la{divergent classical loop}
\parbox[b]{20mm}{\includegraphics[scale=0.4]{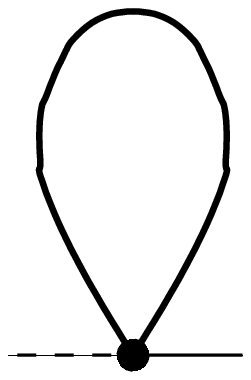}} \!\!\!\!\!\!\!\!\!\! = \, \frac{i g}{2} \int \frac{d^4 p}{(2 \pi)^4} \, \Delta_{\beta} (p) = \frac{i g}{2} \int \frac{d^4 p}{(2 \pi)^4} \,  \frac{2 \pi}{ \beta |p^0|} \, \delta(p^2 - m^2) = \frac{i g}{2 \beta} \int \frac{d^3 p}{(2 \pi)^3} \,\frac{1}{E_{{\bf p}}^2},
\ee
where as usual  $E_{{\bf p}}^2 = {\bf p}^2 + m^2$. This contribution diverges linearly, showing that divergent corrections appear also in the context of classical (thermal) field theory. Notice however that this divergence is milder than the quadratic one encountered in the the case of quantum thermal field theory \cite{Das:1997gg}. On the other hand, it is easy to see that the last two diagrams in equation (\ref{one loop diagrams}) cancel each other, since
\be \la{cancellation}
\parbox[b]{20mm}{\includegraphics[scale=0.4]{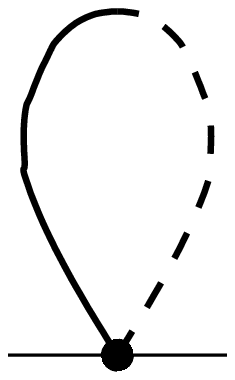}}  \!\!\!\!\!\!\!\!\!\!\! = \,
\frac{i g}{2} \int \frac{d^4 p}{(2 \pi)^4} \, [ - i G_R(p) ] = 
\frac{(-g)}{2} \int \frac{d^4 p}{(2 \pi)^4} \,  [-G_R(p)] =  -  \: \,
\parbox[b]{20mm}{\includegraphics[scale=0.4]{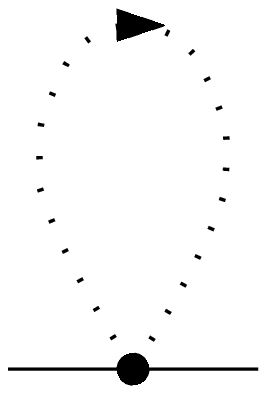}}  \!\!\!\!\!\!\!\!\!\! ,
\ee
where in the last step we have used the fact that a ``Grassmannian'' loop carries an extra minus sign. This cancellation  is a natural consequence of the universal supersymmetry discussed in section \ref{SUSY}. Indeed, similar cancellations  appear also at higher orders in perturbations theory as shown in \cite{Catta10}. Supersymmetry moreover makes it possible to compare the classical and quantum perturbation theory via the introduction of super-diagrams. Further details will appear in \cite{Catta10}. 

Before concluding this subsection, it is worth commenting on a somewhat technical point. The careful reader may think that the  two diagrams in equation (\ref{cancellation}) are both identically zero, since they are proportional to $G_R(x=0)$. However, as already pointed out by Jeon \cite{Jeon:2004dh}, this is not the case unless we choose to regularize the Heaviside step function in such a way that $\theta(t=0)=0$. The regularization of $\theta(t)$ affects in turn the way path integrals are discretized. In particular, the choice $\theta(t=0)=0$ adopted by Aarts and collaborators \cite{Aarts:1996qi, Aarts:1997kp} as well as by other authors \cite{Mueller:2002gd},  corresponds to the so-called \emph{prepoint} (or \emph{Ito}) discretization.

The path integral expression (\ref{eq-13}) for the classical transition amplitude is based instead on the \emph{midpoint} (or \emph{Stratonovich}) discretization which implies the regularization $\theta(0) \equiv \lim_{\epsilon \to 0} [\theta(-\epsilon)+\theta(\epsilon)]/2=1/2$. In fact, a careful calculation of the functional determinant (\ref{eq-11}) in the discretized form shows \cite{Nakazato:1990kk} that the result (\ref{eq-10}) is correct provided the midpoint discretization is adopted. If instead one chooses the prepoint discretization, the functional determinant is just a constant and there is no need to introduce the auxiliary Grassmann variables \cite{Ezawa:1984bm}. However, in the latter case extra terms depending on $\vphi^a$ and $\lambda_a$ may appear in the continuum limit\footnote{A quick calculation of the functional determinant was also carried out in \cite{Gozzi:1989bf}, but that result was not accurate enough to keep the discretization into account.} and perturbation theory may require different diagrams in the discrete and in the continuum limit \cite{Sato:1976hy}. Of course this is not always the case and it depends
on the model under analysis.

It is important to stress that the sum of all one loop corrections (\ref{one loop diagrams}) to the two point function $\langle \phi(x_1) \phi (x_2) \rangle_{\beta}$ does not depend on the way $\theta(t)$ is regularized, by virtue of equation (\ref{cancellation}). Nevertheless, since the Grassmann variables have a physical interpretation as related to the Lyaupunov exponents 
and a geometrical one (see Section \ref{sec:CPI}), plus they play a crucial role in making the universal supersymmetry manifest (see Section \ref{SUSY}), we prefer to adopt the midpoint discretization corresponding to the regularization condition $\theta(0) = 1/2$.

\subsection{Comparison with the CTP formalism at high temperatures} \la{comparison with CTP}

We shall argue now that the approach to classical thermal field theory we just presented \cite{Jeon:2004dh} can be regarded as the \emph{classical} counterpart of the \emph{quantum} CTP formalism. In fact, we will show that, after a suitable change of variables, the propagators in the quantum CTP formalism are approximated by the propagators of classical fields $\phi$ and $\lambda_\pi$ defined in equation (\ref{eq-60})  in the high temperature limit. To this end, let us recall the propagators for a quantum scalar field with a quartic self-interaction \cite{Das:1997gg}, which are characterized by four propagators,
\begin{subequations}
\begin{eqnarray}
G_{1 1}(p) &=& \frac{i}{p^2 - m^2 + i \epsilon} + 2 \pi n_B (|p^0|) \delta(p^2 - m^2)  \\
G_{1 2}(p) &=& 2 \pi \left[ \theta(- p^0) + n_B (|p^0|)\right] \delta(p^2 - m^2) \\
G_{2 1}(p) &=& 2 \pi \left[ \theta( p^0) + n_B (|p^0|)\right] \delta(p^2 - m^2) \\
G_{2 2}(p) &=& - \frac{i}{p^2 - m^2 - i \epsilon} + 2 \pi n_B (|p^0|) \delta(p^2 - m^2), 
\end{eqnarray}
\end{subequations}
where $n_B$ denotes the Bose distribution
\be \la{bose}
n_B (\epsilon) = \frac{1}{e^{\beta \epsilon} - 1}, 
\ee
If we now consider the field redefinition\footnote{Notice that the field redefinition (\ref{fieldridef}) differs from the one commonly adopted in the literature \cite{Aarts:1997kp, Mueller:2002gd, Jeon:2004dh}. This discrepacy can be traced back to the fact that our variables $\lambda_a$ differ by a minus sign from the way analogous variables are usually defined in the literature.}
\begin{equation} \label{fieldridef}
\tilde{\phi} = \left( \begin{array}{c} \tilde{\phi}_1 \\ \tilde{\phi}_2 \end{array}\right) = \left( \begin{array}{cc} 1/2 & 1/2 \\ - 1 & 1 \end{array} \right) \left( \begin{array}{c} \phi_1 \\ \phi_2 \end{array} \right) \equiv U \, \phi,
\end{equation}
the propagators transform as $\tilde{G}_{i j} = \tensor{U}{_i^m} \, \tensor{U}{_j^n} G_{m n}$, and they read:
\begin{eqnarray}
&\tilde{G}_{1 1}(p) =   2 \pi \, [ 1/2 + n_B (|p^0|) ] \,\delta(p^2 - m^2), & \\
& \tilde{G}_{1 2}(p) =  - i G_R (p), \qquad\quad 
\tilde{G}_{2 1}(p) =  - i G_A (p), \qquad \quad
\tilde{G}_{2 2}(p) = 0. \nonumber &
\end{eqnarray}
By comparing these propagators with the classical ones given in equation (\ref{eq-60}) we see that
\begin{equation} \la{otherpropagators}
\tilde{G}_{1 2}(p) = G_{ \phi \lambda_{\pi}}(p), \qquad \tilde{G}_{2 1}(p) = G_{ \lambda_{\pi} \phi}(p), \qquad \tilde{G}_{2 2}(p) = G_{ \lambda_{\pi} \lambda_{\pi}}(p).
\end{equation}
Furthermore, the two sets of propagators become identical in the high temperature limit, since $\tilde{G}_{11}(p)$ reduces to $G_{\phi \phi}(p)$ when $\beta \rightarrow 0$. In fact,
\begin{equation} \la{G11}
\tilde{G}_{1 1} (p) \approx 2 \pi \left( \frac{1}{2} + \frac{1}{\beta |p^0|} \right) \delta(p^2 - m^2) \approx  \frac{2 \pi}{\beta |p^0|} \,\delta(p^2 - m^2) = G_{\phi \phi} (p).
\end{equation}
Thus, equations (\ref{otherpropagators}) and (\ref{G11}) show that, as we anticipated, the free propagators for the quantum CTP formalism are well approximated in the high temperature limit by the classical propagators derived in this section.


\section{Classical Thermofield Dynamics} \la{sec:classicalTFD}

In the previous section we reviewed the approach developed in \cite{Jeon:2004dh}, in which thermal averages are calculated starting from the following expression:
\be \la{starting point Jeon}
\langle \phi(x_1) \, ... \, \phi(x_n) \rangle_{\beta} = Z^{-1}_{\beta} \int [ d \varphi_i] \, \phi_{cl} (x_1; \varphi_i, t_i) \, ... \, \phi_{cl} (x_n; \varphi_i, t_i) \, e^{- \beta H(\varphi_i)}.
\ee
This way of calculating thermal averages is of course very intuitive, since it is based on the idea that the initial conditions of a system in thermal equilibrium must belong to a canonical ensemble.
However, even though we were able to use the Hilbert space formulation of classical mechanics to express equation (\ref{starting point Jeon}) in terms of the expectation value of a time-ordered product of fields, equation (\ref{G-ctp-operatorial}) is not the natural way to calculate statistical averages in the KvN formalism. In fact, the average value of an observable $A(\varphi)$ with respect to a statistical distribution $\rho(\varphi)$ is more naturally given by
\begin{equation} \label{r.5}
\langle A(\varphi)\rangle = \langle \psi | A(\hat{\varphi}) | \psi \rangle,
\end{equation}
where all the information about the statistical ensemble is contained in the state $| \psi \rangle$, which must satisfy the relation $| \langle \psi | \varphi \rangle |^2 = \rho (\varphi)$.

In this section we will introduce an alternative approach to classical thermal field theory in which thermal averages are calculated like in equation (\ref{r.5}). The implementation of such an approach necessarily involves some degree of arbitrariness since, as we pointed out in section \ref{IIa}, a probability distribution $\rho(\vphi)$ can be described by an infinite set of states $|\psi\rangle$ in the KvN space. There is however a very powerful criterium which will allow us to single out a particular state to describe a canonical ensemble. In fact, we will explicitly show that there is only one state in the (extended) KvN space which is invariant under the supersymmetry transformations introduced in section \ref{SUSY}, and that such a state precisely describes a canonical ensemble. 

The idea of expressing thermal averages of observable quantities as expectation values of operators acting on a physically redundant Hilbert space is also the starting point of an alternative approach to \emph{quantum} thermal field theory known as Thermofield Dynamics \cite{Umezawa:1982nv}. As we will see, the correspondence between Thermofield Dynamics and the classical approach we will develop in this section is not only qualitative, but also quantitative. In fact, we will derive the classical Feynman rules for a scalar field within this alternative classical approach, and we will show that the propagators for $\phi$ and $\lambda_\pi$ perfectly match the quantum propagators of Thermofield Dynamics in the high temperature limit.

\subsection{Canonical ensemble and invariance under supersymmetry} \label{par:stato}

We will now show that the (extended) KvN Hilbert space admits only one supersymmetry-invariant state, and that this state is associated with the canonical ensemble. For simplicity, we will carry out our analysis for a system with a finite number of degrees of freedom, but the result can be  extended to field theories in a very straightforward way. Let us start by rewriting the supersymmetry charges introduced in section \ref{SUSY}:
\begin{equation} \label{x.4}
\begin{array}{l} \hat{Q}_H  = \hat{c}^a \left( i \hat{\lambda}_a + \beta \partial_a H(\hat{\varphi})/2 \right) \vspace{0.3cm} \\
\hat{\bar{Q}}_H  = \hat{\bar{c}}_a \omega^{a b} \left( i \hat{\lambda}_b - \beta \partial_b H(\hat{\varphi})/2 \right). \end{array}
\end{equation}
where $\beta$ is a real parameter. The request that a state  $| \psi \rangle$ associated with a probability distribution $\rho (\varphi)$ be invariant under supersymmetry can be expressed in the following way
\begin{equation} \label{t.4}
\hat{Q}_H | \psi \rangle = \hat{\bar{Q}}_H | \psi \rangle = 0.
\end{equation}

In order to explore the restrictions that these conditions impose on the state $| \psi \rangle$, let us consider the associated wave function in the $(\vphi, c)$ space \cite{Deotto:2002hy}:
\begin{equation} \label{r.6}
\psi(\varphi, c) \equiv \langle (-i c^{p})^* +, \left(i c^{q}\right)^* +, \varphi | \psi \rangle \equiv \psi_0 (\varphi) + \psi_q (\varphi) \, c^q + \psi_p (\varphi) \, c^p + \psi_2 (\varphi) \, c^q c^p.
\end{equation}
Notice that, according to the notation introduced in appendix \ref{app:symplectic} and due to the Hermiticity property $(\hat{c}^a)^\dag = i \omega^{a b} \hat{\bar{c}}_b$, the bra $\langle (-i c^{p})^* +, \left(i c^{q}\right)^* +, \varphi |$ is a simultaneous eigenstate of $\hat{\vphi}^a$ and $\hat{c}^a$ with eigenvalues $\vphi^a$ and $c^a$. In this representation the operators  $\hat{\lambda}_a$ and  $\hat{\bar{c}}_a$ can be implemented as derivative operators, such that
\ba
\langle (-i c^{p})^* +, \left(i c^{q}\right)^* +, \varphi | \hat{\lambda}_a | \psi \rangle &=&  - i \frac{\partial}{\partial \vphi^a} \, \psi(\varphi, c) \\
\langle (-i c^{p})^* +, \left(i c^{q}\right)^* +, \varphi | \hat{\bar{c}}_a | \psi \rangle &=& \frac{\partial}{\partial c^a} \, \psi(\varphi, c).
\ea
As a consequence the request $\hat{Q}_H | \psi \rangle =0$ can be rewritten as follows:
\begin{eqnarray}
0 &=& \langle (-i c^{p})^* +, \left(i c^{q}\right)^* +, \varphi | \hat{Q}_H | \psi \rangle \nonumber \\
&=&  \langle (-i c^{p})^* +, \left(i c^{q}\right)^* +, \varphi | \hat{c}^a \left( i \hat{\lambda}_a + \beta \partial_a H(\hat{\varphi})/2 \right) | \psi \rangle \la{0=Q_H psi} \\
&=& c^a \left( \partial_a +  \beta \partial_a H(\varphi)/2 \right) \psi (\varphi, c), \nonumber
\end{eqnarray}
If we now use the explicit form (\ref{r.6}) for $\psi (\varphi, c)$, we see that equation (\ref{0=Q_H psi}) is equivalent to the following two relations:
\begin{eqnarray}
&\left( \partial_a +  \beta \partial_a H(\varphi)/2 \right)  \psi_0 (\varphi) = 0 & \label{r.8} \\
&\left( \partial_p +  \beta \partial_p H(\varphi)/2 \right) \psi_q (\varphi) = \left( \partial_q +  \beta \partial_q H(\varphi)/2 \right) \psi_p (\varphi). &\label{r.7}
\end{eqnarray}
The first equation can be easily solved for any form of the Hamiltonian $H$, and it has the following solution:
\begin{equation} \la{psi0}
\psi_0 (\varphi) = K_0 \, e^{- \beta H (\varphi) /2},
\end{equation}
where $K_0$ is for now an arbitrary integration constant. If we now assume that in the  Hamiltonian of the system $p$ and $q$ do not couple to each other, it is then reasonable to make the assumption that $\psi_q $ and $\psi_p $ have a factorized form $\psi_q (\varphi) = f_q (q) \, g_q (p)$ and $\psi_p (\varphi) = f_p (q) \, g_p (p)$. Under these assumption, equation (\ref{r.7}) is then equivalent to the following set of equations:
\begin{eqnarray}
\left( \partial_p +  \beta \partial_p H(\varphi)/2 \right) g_q (p) &=& g_p (p) \label{r.9}\\
\left( \partial_q +  \beta \partial_q H(\varphi)/2 \right) f_p (q) &=& f_q (q). \label{r.0}
\end{eqnarray}

Let us now turn to  the second condition   that $|\psi\rangle$ must satisfy in order  to be invariant under supersymmetry, namely $\hat{\bar{Q}}_H | \psi \rangle = 0$. This condition can be expressed in terms of the wave function $\psi(\vphi,c)$ as 
\begin{eqnarray}
0 &=& \langle (-i c^{p})^* +, \left(i c^{q}\right)^* +, \varphi | \hat{\bar{Q}}_H | \psi \rangle \nonumber \\
&=&  \langle (-i c^{p})^* +, \left(i c^{q}\right)^* +, \varphi | \hat{\bar{c}}_a \omega^{a b} \left( i \hat{\lambda}_b - \beta \partial_b H(\hat{\varphi})/2 \right) | \psi \rangle \\
&=& \frac{\partial}{\partial c^a}  \omega^{a b} \left( \partial_b -\beta \partial_b H(\varphi)/2 \right) \psi(\varphi, c), \nonumber
\end{eqnarray}
or, by using equation (\ref{r.6}), as
\begin{eqnarray}
&\left( \partial_a -  \beta \partial_a H(\varphi)/2 \right)  \psi_2 (\varphi) = 0 &\\
&\left( \partial_p -  \beta \partial_p H(\varphi)/2 \right) \psi_q (\varphi) = \left( \partial_q -  \beta \partial_q H(\varphi)/2 \right) \psi_p (\varphi). & \label{b5}
\end{eqnarray}
The first equation can be solved analogously to (\ref{r.8}) and yields
\begin{equation}
\psi_2 (\varphi) = K_2 \, e^{ \beta H (\varphi) /2},
\end{equation}
where $K_2$ is for now another arbitrary constant. Next, by using the factorized form of $\psi_q$ and $\psi_p$, we can trade equation (\ref{b5}) for the following two equations:
\begin{eqnarray}
\left( \partial_p -  \beta \partial_p H(\varphi)/2 \right) g_q (p) &=& g_p (p) \label{s.1} \\
\left( \partial_q -  \beta \partial_q H(\varphi)/2 \right) f_p (q) &=& f_q (q). \label{s.2}
\end{eqnarray}
These two equations, together with equations (\ref{r.9}) and (\ref{r.0}), indicate that both $\psi_q$ and $\psi_p$ must vanish. In fact, by subtracting equation (\ref{s.1}) from equation (\ref{r.9}) we get
\begin{equation}
 \beta \partial_p H(\varphi) g_q (p) = 0 \quad \Longrightarrow \quad g_q (p) = 0 \quad \Longrightarrow \quad \psi_q (\varphi) = 0,
\end{equation}
and similarly, from the difference between equations (\ref{s.2}) and (\ref{r.0}) we derive that
\begin{equation}
 \beta \partial_q H(\varphi) f_p (q) = 0 \quad \Longrightarrow \quad f_p (q) = 0 \quad \Longrightarrow \quad \psi_p (\varphi) = 0.
\end{equation}

To summarize, what we have shown so far is that invariance under supersymmetry requires the state $\psi (\varphi, c)$ to have the following form
\begin{equation} \label{s.5}
\psi (\varphi, c) = K_0 \, e^{-\beta H (\varphi) /2} + K_2 \, e^{ \beta H (\varphi) /2} \, c^q c^p.
\end{equation}
We will now see that the integration constants $K_0$ e $K_2$ can be determined by considering the probability distribution associated with a state of the form (\ref{s.5}). An instructive way to do that is to rewrite the expectation value of an observable $A(\varphi)$ on the state $| \psi \rangle$ by using the completeness relations discussed in the appendix \ref{app:symplectic}:
\be\la{expect value with completeness}
\langle A(\varphi)\rangle = \langle \psi | A(\hat{\varphi}) | \psi \rangle = \int d\varphi \, d c^q d c^p A(\varphi) \langle \psi | \varphi, c^q -, c^p - \rangle \langle (-i c^{p})^* +, \left(i c^{q}\right)^* +, \varphi | \psi \rangle.
\ee
Clearly, in order for equation (\ref{expect value with completeness}) to agree with the usual definition of average value with respect to the probability density $\rho(\varphi)$,  the following relation must hold true:
\begin{equation} \label{s.9}
\int d c^q d c^p \langle \psi | \varphi, c^q -, c^p - \rangle \langle (-i c^{p})^* +, \left(i c^{q}\right)^* +, \varphi |\psi \rangle = \rho(\varphi).
\end{equation}
The integrals on the LHS of this equation can be easily calculated by using the fact that \break
 $\langle i c^{p *} +, \left(- i c^{q *}\right) +, \varphi | \psi \rangle = \psi (\varphi, c)$ is given by (\ref{s.5}) and that
\begin{eqnarray}
\langle  c^q -,  c^p -, \varphi | \psi \rangle &=& \int d \varphi^{\prime} \, d c^{q \prime} d c^{p \prime} \langle  c^q -,  c^p -, \varphi | \varphi^{\prime}, c^{q \prime} -, c^{p \prime} -\rangle  \langle i c^{p \prime *} +, \left(- i c^{q \prime  *}\right) +, \varphi^{\prime} | \psi \rangle \nonumber \\
&=& - \int d c^{q \prime} d c^{p \prime} \exp \left( - i c^{q *} c^{p \prime} + i c^{p *} c^{q \prime} \right) \psi (\varphi, c^{\prime}) \nonumber \\
&=& -  K_2 \, e^{ \beta H (\varphi) /2} +  K_0 \, e^{ - \beta H (\varphi) /2} \, c^{p *} c^{q *}. \label{t.5}
\end{eqnarray}
Equation (\ref{t.5}) can be easily proved by using the results we derived in appendix \ref{app:symplectic}, and can be combined with equation (\ref{s.9}) to get:
\be
| K_0 |^2 e^{- \beta H(\varphi)} - | K_2 |^2 e^{\beta H(\varphi)} = \rho(\varphi),
\ee
This result shows that, unless the Hamiltonian is bounded from above or $\beta=0$, the only way to have a probability distribution $\rho(\vphi)$ which is positive definite and can be appropriately normalized is to have $K_2=0$ and $\beta >0$. Then, the value of $K_0$ is fixed by the request $\rho (\varphi)$ is normalized to one, and the final result is that the only supersymmetric invariant state in the extended Hilbert space which corresponds to a physical probability distribution is associated with the canonical ensemble:
\begin{equation} \label{t.1}
\psi_{\beta} (\varphi, c) = Z^{ - 1/2}_{\beta} e^{- \beta H(\varphi) / 2},
\end{equation}
where the parameter $\beta$ can now be interpreted as the inverse temperature. Notice that our analysis does not depend on the number of degrees of freedom described by the Hamiltonian $H(\varphi)$, and therefore this result is also valid in the case of classical field theories as long as  there is no coupling between configuration and momentum variables. 

Thus, the fact that essentially the only state which is invariant under supersymmetry is associated with the canonical ensemble allows us to interpret the requirement of supersymmetry invariance as an algebraic characterization of a system in thermal equilibrium \cite{Gozzi:1989xz}.

\subsection{Generating functional and Feynman rules }

We are now going to use the state (\ref{t.1}) to develop an alternative approach to classical thermal field theory, in which thermal correlation functions of a scalar field $\phi$ are calculated according to the formula
\begin{equation}
\langle \phi(x_1) \, ... \, \phi(x_n) \rangle_{\beta} = \langle \psi_{\beta} | \, T [ \hat{\phi}(x_1) \, ... \, \hat{\phi}(x_n) ] \,  | \psi_{\beta} \rangle,
\end{equation}
where  $| \psi_{\beta} \rangle$ is the supersymmetric invariant state derived in the previous section. Clearly, the correlation functions above can be obtained by deriving the generating functional  
\begin{equation} \label{b8}
Z \equiv \langle \psi_{\beta} | \, T \, \exp \left\{ i \int_{t_i}^{t_f} dt \, d^3 x \left( J_{\phi} \, \hat{\phi} + \Lambda^{\pi} \hat{\lambda}_{\pi} - i \bar{\eta}_{\phi} \hat{c}^{\phi} - i \hat{\bar{c}}_{\pi} \eta^{\pi} \right) \right\} \, | \psi_{\beta} \rangle
\end{equation}
with respect to the current $J_\phi$ as in equation (\ref{derivatives wrt Jphi}). As in the case of the classical CTP approach, the currents coupled to $\lambda_\pi, c^\phi$ and $\bar{c}_\pi$ are necessary to set up perturbative calculations. Notice also that, for now, we are considering some finite initial and final times $t_i$ e $t_f$, but eventually we will take the limits $ t_i \to - \infty$  and  $t_f \to + \infty$. By inserting two completeness relations at the instants $t_i$ and $t_f$, we can rewrite  $Z$  the following way:
\begin{eqnarray}
Z &=& \int [d \varphi_f] [d c_f] [d \varphi_i] [d c_i] \, \langle  \psi_{\beta}| \varphi_f, c^\phi_f -, c^\pi_f -, t_f \rangle \, \times \label{t.8} \\
&& \times \, \langle i c^{\pi *}_f +, (- i c^{\phi *}_f ) +, \varphi_f, t_f | T[...] | \varphi_i, c^{\phi}_i -, c^{\pi}_i -, t_i \rangle \langle i c^{\pi *}_i +, (- i c^{\phi *}_i ) +, \varphi_i, t_i  | \psi_{\beta} \rangle. \nonumber
\end{eqnarray}
Each of the three elements in the integrand can be calculated in a straightforward way. Let us start by noticing that the state  $| \psi_{\beta} \rangle$ is stationary, because it is an eigenstate of $\hat{\mathcal{H}}$ with eigenvalue $0$. In fact, this can be easily shown by using equation (\ref{eq-t.3}) to express $\hat{\mathcal{H}}$ in terms of the supersymmetry charges:
\be
\hat{\mathcal{H}} | \psi_{\beta} \rangle = \frac{i}{2 \beta} \, \left\{ \hat{Q}_H, \hat{\bar{Q}}_H \right\} | \psi_{\beta} \rangle = 0.
\ee
Then, the first term in the integrand in equation (\ref{t.8}) can be easily obtained by taking the complex conjugate of equation (\ref{t.5}) with $K_2=0$ and $K_0 = Z_{\beta}^{-1/2}$, and using the fact that $|\psi_\beta\rangle$ does not evolve in time:
\begin{eqnarray}
\langle \psi_{\beta} | \varphi_f, c^\phi_f -, c^\pi_f -, t_f \rangle &=& \langle  \psi_{\beta} | e^{ i \hat{\mathcal{H}} t_f} | \varphi_f, c^\phi_f -, c^\pi_f - \rangle \\
&=& \langle  \psi_{\beta} | \varphi_f, c^\phi_f -, c^\pi_f - \rangle =  Z^{ - 1/2}_{\beta} e^{- \beta H(\varphi_f) / 2} \, c^\phi_f c^\pi_f. \nonumber
\end{eqnarray}
The last term in the integrand is by definition $ \psi_{\beta} (\varphi_i, c_i)$ and therefore, according to equation (\ref{t.1}), we have
\be
\langle i c^{\pi *}_i +, (- i c^{\phi *}_i ) +, \varphi_i, t_i  | \psi_{\beta} \rangle = Z^{ - 1/2}_{\beta} e^{- \beta H(\varphi_i) / 2}.
\ee
Finally, the second term in the integrand in equation (\ref{t.8}) can be calculated using the standard slicing procedure together with the path integral expression for the transition amplitude (\ref{eq-13}). Therefore, an opportunely normalized path integral expression for $Z$ reads:
\be\la{CTFD's generating functional}
Z \equiv  \frac{ \displaystyle \int \mathcal{D} \mu \,\exp \left\{ i \, \mathcal{S} - \fr{\beta}{2}[ H(\vphi(t_i)) + H(\vphi(t_f))]  + i \int \l( \Lambda^\pi \lambda_\pi + J_\phi \phi - i \bar{\eta}_\phi c^\phi - i \bar{c}_\pi \eta^\pi\r) \right\}}{ \displaystyle \int \mathcal{D} \mu \, \exp \left\{ i \, \mathcal{S} - \fr{\beta}{2}[ H(\vphi(t_i)) + H(\vphi(t_f))] \right\}}
\ee
where ${\mathcal D}\mu\equiv\mathcal{D} \varphi \, \mathcal{D} \lambda \, \mathcal{D} c\, \mathcal{D} \bar{c}$, and $\mathcal{S}$ is given by the sum of the three terms in equations (\ref{S1}), (\ref{S2}) and (\ref{S3}). Notice that the factor $c^\phi_f c^\pi_f$ was absorbed in the measure of integration, which is now equal to the one used in the previous section. However, the generating functional (\ref{CTFD's generating functional}) is clearly different from the one of the classical CTP formalism (see equation (\ref{Jeon's generating functional})) and as such it will give rise to a different set of Feynman rules.

Notice that the only difference between equations (\ref{Jeon's generating functional}) and (\ref{CTFD's generating functional}) consists in the weights associated with the initial conditions $\varphi_i$ to the final ones $\varphi_f$.  For this reason, the Feynman rules for the propagators of the Grassmann variables will be equal to the ones we derived for the classical CTP formalism, see equation (\ref{eq-60}) and Appendix \ref{app: ghost propagator}. Furthermore, in the limit where $ t_i \to - \infty$ and $t_f \to \infty$, also the Feynman rules for the vertices remains the same as in equation (\ref{classical CTP vertices}), since we are assuming that the interaction is turned off adiabatically in the far past and future. Thus, the only Feynman rules that are different with respect to the ones derived in the previous section are those for the propagators of $\phi$ and $\lambda_\pi$, which follow from the ``reduced'' generating functional
\be \la{reduced generating functional CTFD}
Z_1[J_\phi, \Lambda^\pi] = \frac{\displaystyle\int \mathcal{D} \varphi \, \mathcal{D} \lambda \, \exp \left\{  i \, \mathcal{S}_1 - \fr{\beta}{2}[ H(\vphi(t_i)) + H(\vphi(t_f))]   + i \int \l( \Lambda^\pi \lambda_\pi + J_\phi \phi \r) \right\}}{\displaystyle \int \mathcal{D} \varphi \, \mathcal{D} \lambda \,\exp \left\{ i \, \mathcal{S}_1 - \fr{\beta}{2}[ H(\vphi(t_i)) + H(\vphi(t_f))]  \right\}} .
\ee
This generating functional can be calculated by using the result\footnote{Notice that equation (\ref{penco mauro result}) can also be derived from (\ref{penco mauro original result}) by simply integrating over $\vphi_f$.}
\ba
&& \int \mathcal{D}'' \varphi \, \mathcal{D} \lambda \, \exp \left\{  i \, \mathcal{S}_1 + i \int \l( \Lambda^\pi \lambda_\pi + J_\phi \phi \r) \right\} = \la{penco mauro original result}\\
&& \qquad\qquad\qquad\qquad\qquad = \mathcal{N} \, \delta (\vphi_f - \tilde{\vphi}_{cl}(t_f ; \vphi_i, t_i)) \exp \l\{ i \int_{t_i}^{t_f} d^4 x \, J_\phi \, \tilde{\phi}_{cl}\r\} , \nonumber
\ea
where $\tilde{\vphi}_{cl}$ are the classical solutions to the equations of motion associated with the modified Hamiltonian $\tilde{H} = H + \int \Lambda^\pi \phi$ \cite{Penco:2006wn}. Then, the integration over $\vphi_f$ in (\ref{reduced generating functional CTFD}) becomes trivial, while the one over the initial conditions $\vphi_i$ is just a Gaussian integral. The final result is the following: 
\begin{eqnarray}
Z_1 [J_{\phi}, \Lambda^\pi] &=& \exp \left\{ - \frac{1}{2}  \int d^4 x \, d^4 x^{\prime} \left[  J_{\phi} (x)G_{\phi \phi} (x - x^{\prime}) J_{\phi} (x^{\prime}) \, + \right. \right. \nonumber \\
&& \left. \qquad \qquad + \, 2 J_{\phi} (x) G_{\phi \lambda_\pi} (x - x^{\prime}) \Lambda^\pi (x^{\prime}) + \Lambda^\pi (x) G_{\lambda_\pi \lambda_\pi} (x - x^{\prime}) \Lambda^\pi (x^{\prime}) \right] \Biggr\},  \nonumber 
\end{eqnarray}
with 
\begin{eqnarray} \la{CTFD propagators}
&& \!\!\!\! \displaystyle G_{\phi \phi} = \, \parbox{20mm}{
\begin{fmfgraph}(65,30) 
\fmfleft{i1}
\fmfright{o1}
\fmf{phantom}{i1,o1}
\fmffreeze
\fmf{vanilla}{i1,o1}
\fmfv{decor.shape=circle, decor.filled=full, decor.size=1.5thick}{i1,o1}
\end{fmfgraph}} \quad = \Delta_{\beta} (p), \qquad \,\,\,\,
G_{ \phi \lambda_\pi} = \, \parbox{20mm}{
\begin{fmfgraph}(65,30) 
\fmfleft{i1}
\fmfright{o1}
\fmf{phantom}{i1,v,o1}
\fmffreeze
\fmf{vanilla}{i1,v}
\fmf{dashes}{v,o1}
\fmfv{decor.shape=circle, decor.filled=full, decor.size=1.5thick}{i1,o1}
\end{fmfgraph}} \quad = - \frac{i}{2} \left[ G_R (p) + G_A (p) \right]  \\
&& \!\!\!\! \displaystyle G_{ \lambda_\pi \phi} = \, \parbox{20mm}{
\begin{fmfgraph}(65,30) 
\fmfleft{i1}
\fmfright{o1}
\fmf{phantom}{i1,v,o1}
\fmffreeze
\fmf{vanilla}{v,o1}
\fmf{dashes}{i1,v}
\fmfv{decor.shape=circle, decor.filled=full, decor.size=1.5thick}{i1,o1}
\end{fmfgraph}} \quad = - \frac{i}{2} \left[ G_R (p) + G_A (p) \right] , \quad
G_{ \lambda_\pi \lambda_\pi}  = \, \parbox{20mm}{
\begin{fmfgraph}(65,30) 
\fmfleft{i1}
\fmfright{o1}
\fmf{dashes}{i1,o1}
\fmfv{decor.shape=circle, decor.filled=full, decor.size=1.5thick}{i1,o1}
\end{fmfgraph}} \quad = \frac{\beta^2 (p^0)^2}{4} \, \Delta_{\beta} (p)  \nonumber
\end{eqnarray}
and $\Delta_\beta$ defined as in equation (\ref{delta beta}). These Feynman rules are different from the ones derived in the previous section and displayed in equation (\ref{eq-60}). Notice however that the propagator $G_{\phi \phi}$ is still the same. Clearly this had to be the case, since $G_{\phi \phi} (x - x^{\prime})$ is just the average value $\langle \phi (x) \phi(x^{\prime}) \rangle_{\beta}$ calculated in the free case and, as such, it is a physical quantity that cannot depend on the way thermal averages are implemented. On the other hand, there is no reason why the Feynman rules for the other propagators should also remain the same, since $\lambda_\pi$ is not an observable quantity. In particular, the most striking difference between the Feynman rules in (\ref{eq-60}) and the ones we just derived is that the propagator $G_{ \lambda_\pi \lambda_\pi}$ is no longer zero.

\subsection{Equivalence to the classical CTP formalism} \la{eqivCTP}

In order to convince ourselves that the approach to classical thermal field theory we just developed is physically equivalent to the classical CTP approach, we will now reconsider the 1-loop correction to the 2-point function $\langle \phi \phi \rangle_\beta$, given by the sum of the following diagrams:
\begin{eqnarray}
\parbox[b]{20mm}{\includegraphics[scale=0.4]{1loop3.eps}} \quad + \quad  \parbox[b]{20mm}{\includegraphics[scale=0.4]{1loop4.eps}} \quad &+& \quad
\parbox[b]{20mm}{\includegraphics[scale=0.4]{1loop1.eps}} \quad + \quad
\parbox[b]{20mm}{\includegraphics[scale=0.4]{1loop2.eps}} \ .  \nonumber \\
 \qquad (A) \qquad \: \, \: \quad \: \, \: \: \, \qquad (B) \: \: \: \qquad && \: \: \: \qquad (C) \: \: \, \qquad \: \: \, \: \quad \qquad (D) \nonumber
\end{eqnarray}
If we denote by $\Sigma_{\beta}(m^2)$ the loop made out of $G_{\phi \phi}$, which we already calculated in equation (\ref{divergent classical loop}), then we can easily calculate the sum of the first two diagrams by using the Feynman rules we just derived:
\begin{eqnarray}
(A) + (B) &=& \Delta_{\beta} (p) \Sigma_{\beta}(m^2) \left( - \frac{i}{2} [ G_A (p) + G_R (p)] \right) -  \frac{i}{2} [ G_A (p) + G_R (p)] \Sigma_{\beta}(m^2) \Delta_{\beta} (p) \nonumber \\
&=&- i \Delta_{\beta} (p) [ G_R (p) + G_A (p)]  \Sigma_{\beta}(m^2). 
\end{eqnarray}
This result agrees with the one we would obtain by using the Feynman rules (\ref{eq-60}) and (\ref{classical CTP vertices}) for the classical CTP approach:
\begin{eqnarray}
(A) + (B) &=& \Delta_{\beta} (p) \Sigma_{\beta}(m^2) [ - i G_A (p)] + [ - i G_R (p)] \Sigma_{\beta}(m^2) \Delta_{\beta} (p) \nonumber \\
&=& - i \Delta_{\beta} (p) [ G_R (p) + G_A (p)]  \Sigma_{\beta}(m^2) . \la{1loopclasrealtime}
\end{eqnarray}
Moreover, it is easy to see that the remaining diagrams $(C)$ and $(D)$, cancel each other also with this new set of Feynman rule, since the relation (\ref{cancellation}) still holds true:
\begin{eqnarray}
\parbox[b]{20mm}{\includegraphics[scale=0.4]{onlyloop2.eps}}  \!\!\!\!\!\!\!\!\!\!\! &=& \,
\frac{i g}{2} \int \frac{d^4 p}{(2 \pi)^4} \, \left( - \frac{i}{2} [ G_A (p) + G_R (p)] \right)  \\
&=& \frac{g}{4} \left\{ \int \frac{d^4 p}{(2 \pi)^4} \, G_R (-p) + \int \frac{d^4 p}{(2 \pi)^4} \, G_R (p)\right\} \nonumber = - \frac{g}{2} \int \frac{d^4 p}{(2 \pi)^4} \, [ - G_R(p) ] = - \,  \parbox[b]{20mm}{\includegraphics[scale=0.4]{onlyloop3.eps}} \!\!\!\!\!\!\!\!\!\! .
\end{eqnarray}
This simple calculation illustrates how the formalism developed in this section yields the same physical results as the classical CTP formalism \cite{Jeon:2004dh}, even though the Feynman rules are slightly different in the two cases.


\subsection{Comparison with the Thermofield Dynamics at high temperatures} \label{highTFD}

We have seen in sections \ref{comparison with CTP} that, at high temperatures, there is a very important connection between classical and quantum thermal field theories. In fact, after a suitable change of variables, the free propagators of the {\it quantum} CTP formalism can be approximated by the \emph{classical} propagators (\ref{eq-60}) for $\phi$ and $\lambda_\pi$ in the limit $\beta \rightarrow 0$.

However, the CTP formalism is only one of the possible implementations of quantum thermal field theory, as reviewed in Section \ref{sec:QTFT}. For this reason, one might wonder whether the alternative formulation of classical thermal field theory we introduced in this section is related to any known quantum formalism. We will now show that this is the case, and that the propagators (\ref{CTFD propagators}) are connected to the propagators of quantum Thermofield Dynamics  \cite{Umezawa:1982nv} in the limit $\beta \rightarrow 0$.

Let us start then by summarizing the Feynman rules for the TFD propagators, which can be easily obtained from equations (\ref{realtimeformalismspropagators}) with $\sigma =1/2$: 
\begin{subequations}
\begin{eqnarray}
G_{1 1}(p) &=& \frac{i}{p^2 - m^2 + i \epsilon} + 2 \pi n_B (|p^0|) \delta(p^2 - m^2) \\
G_{1 2}(p) &=& 2 \pi \, e^{ \beta  |p^0| /2} \, n_B (|p^0|) \delta(p^2 - m^2)  \\
G_{2 1}(p) &=& 2 \pi \, e^{ \beta  |p^0| /2} \, n_B (|p^0|) \delta(p^2 - m^2) \\
G_{2 2}(p) &=& - \frac{i}{p^2 - m^2 - i \epsilon} + 2 \pi n_B (|p^0|) \delta(p^2 - m^2). 
\end{eqnarray}
\end{subequations}
Once again, $n_B$ stands for the Bose distribution given in equation (\ref{bose}). 
If we now introduce the same change of variables we considered in section \ref{comparison with CTP}, namely 
\begin{equation} \label{r.2}
\tilde{\phi} = \left( \begin{array}{c} \tilde{\phi}_1 \\ \tilde{\phi}_2 \end{array}\right) = \left( \begin{array}{cc} 1/2 & 1/2 \\ - 1 & 1 \end{array} \right) \left( \begin{array}{c} \phi_1 \\ \phi_2 \end{array} \right) \equiv U \, \phi .
\end{equation}
the propagators  $G_{i j}$ of the TFD approach transform according to $\tilde{G}_{i j} = \tensor{U}{_i^m} \, \tensor{U}{_j^n} G_{m n}$ into:
\begin{eqnarray}
\tilde{G}_{1 1}(p) &=& \frac{\pi}{2} \, n_B (|p^0|) \left( e^{\beta | p^0| / 2} + 1 \right)^2 \delta(p^2 - m^2) \nonumber \\
\tilde{G}_{1 2}(p) &=& - \frac{i}{2} [ G_R (p) + G_A (p) ] \\
\tilde{G}_{2 1}(p) &=& - \frac{i}{2} [ G_R (p) + G_A (p) ] \nonumber \\
\tilde{G}_{2 2}(p) &=& 2 \pi \, n_B (|p^0|) \left( e^{\beta | p^0| / 2} - 1 \right)^2 \delta(p^2 - m^2). \nonumber
\end{eqnarray}
Notice that the mixed propagators  are independent of the temperature and are equal to the corresponding classical ones :
\begin{equation}
\tilde{G}_{1 2}(p) =  G_{\phi \lambda_\pi}(p), \qquad \tilde{G}_{2 1}(p) = G_{\lambda_\pi \phi}(p).
\end{equation}
The other two propagators can instead be linked to the classical ones (\ref{CTFD propagators}) in the limit $\beta \rightarrow 0$:
\begin{eqnarray}
\bar{G}_{1 1}(p) &\approx& \frac{\pi}{2} \, \frac{1}{ \beta | p^0 |}  \left( 1 + 1 \right)^2 \delta(p^2 - m^2) = \frac{2 \pi}{\beta | p^0 |} \, \delta(p^2 - m^2) = G_{\phi \phi}(p) \nonumber \\
\bar{G}_{2 2}(p) &\approx& 2 \pi \, \frac{1}{ \beta | p^0 |}  \left( \frac{\beta |p^0|}{2} \right)^2 \delta(p^2 - m^2) = \frac{ \pi \beta | p^0 | }{2} \, \delta(p^2 - m^2) = G_{\lambda_\pi \lambda_\pi}(p). \nonumber
\end{eqnarray}
Thus, we have shown that the Thermofield Dynamics formalism is the quantum counterpart to the classical formalism developed in this section, which will therefore be called \emph{classical Thermofield Dynamics}.

In conclusion, it is interesting to point out that the relation between classical and quantum Thermofield Dynamics is not only limited to the propagators, and it has a far deeper conceptual nature. In fact, quantum Thermofield Dynamics was originally conceived as an operatorial approach in which thermal averages of quantum fields are calculated as expectation values over the so called thermal vacuum, a state belonging to an enlarged Hilbert space and such that
\begin{equation}
\langle A \rangle_{\beta} = \langle 0, \beta | \hat{A} | 0, \beta \rangle.
\end{equation}
Clearly, the ideas behind the quantum Thermofield Dynamics approach bear a close similarity to the KvN approach to classical mechanics we relied upon throughout this section. From this point of view, the existence of a connection between classical and quantum propagators like the one we outlined above is not very surprising after all.


\section{Classical Matsubara formalism} \la{sec:classicalMatsubara}

In the previous  two sections we introduced two different approaches to classical thermal field theory which, at high temperatures, turned out to be related to two well know real time formulations of  quantum thermal field theory, namely the CTP formalism and the Thermofield Dynamics. This classical-quantum connection was possible because, in both cases, the number of variables needed to set up a path integral formulation were twice the number of physical degrees of freedom (without considering the Grassmann variables $c^a$ and~$\bar{c}_a$).

However, in section \ref{sec:QTFT} we reviewed a third approach to quantum thermal field theory, namely the Matsubara formalism \cite{Matsubara:1955ws}. This approach is less powerful than the real-time formalisms, since it can be used almost exclusively\footnote{Analytical continuations are possible, although they often involve technical subtleties.} to calculate properties of a system in thermal equilibrium, but its formulation does not require the introduction of additional degrees of freedom. It is then natural to wonder whether it is possible to introduce a third formulation of classical thermal field theory which is connected to the quantum Matsubara formalism at high temperatures. As we will see in this section, such a formulation exists and shares many formal aspects with its quantum counterpart, namely:
\begin{itemize}
\item it is a ``trace formalism'';
\item thermal averages are implemented via the statistical operator $\hat{\rho} = e^{- \beta H(\hat{\vphi})}$;
\item the statistical operator $\hat{\rho}$ can be interpreted as a time evolution operator in a suitably defined imaginary direction (more on this later).
\end{itemize}
A further and very important feature of this new approach is that it relies heavily on the superfield formulation introduced in section \ref{SUSY}, and as such
\begin{itemize}
\item it does not require (formally) a doubling of  ``variables'', since it works directly with the supermultiplets $\Phi^a$ instead of dealing with the variables $\vphi^a$ and $\lambda_a$ separately. 
\end{itemize}
For all these reasons, we will name this formalism \emph{classical Matsubara approach}.

\subsection{Statistical operator and imaginary time evolution}

As we already pointed out in section \ref{IIa}, the KvN formulation of classical mechanics is physically redundant, since every statistical distribution in phase space is associated with an infinite set of states in the KvN space. For this reason, although expectation values were originally  defined as $\langle O(\vphi) \rangle_\rho = \langle \psi | \hat{O} | \psi \rangle$ with $| \psi \rangle$ one of the infinite number of states associated with the probability distribution $\rho(\vphi)$, we could also calculate them as follows:
\begin{equation} 
\langle O(\vphi) \rangle_\rho = \fr{\mbox{Tr}[\hat{\rho} \, \hat{O}]}{\mbox{Tr}\hat{\rho}} \ , \qquad \qquad \quad \hat{\rho} \equiv \sum_{\alpha} f_{\alpha} |\psi_{\alpha} \rangle \langle \psi_{\alpha} |,
\end{equation}
where the states $|\psi_{\alpha} \rangle$ are all associated with the probability distribution $\rho(\vphi)$, and the coefficients $f_{\alpha}$ are real. This redundancy in the description of a physical system is ultimately responsible for the existence of several equivalent approaches to classical thermal field theory.

In this section, we will explore an approach based on the Gibbs statistical operator: 
\be \la{Gibbs}
\hat{\rho} = e^{- \beta H(\hat{\vphi})}.
\ee
The fact that this is the same statistical operator used in quantum thermal field theory is very intriguing. However, it is important to stress that the properties of the statistical operator (\ref{Gibbs}) depend crucially on the Hilbert space it acts upon, and since the Hilbert spaces of classical and quantum mechanics are different, classical and quantum correlation functions will in general be different even though they are calculated by using the same statistical operator. A simple illustration of this fact is provided by the trace of the statistical operator (\ref{Gibbs}), which is generally finite for a quantum system, but it is divergent\footnote{Dealing with statistical operators with divergent trace requires particular attention. For our purposes, the most relevant difficulty one encounters is that the cyclicity of the trace may fail. This will not be a problem in this section since the approach we will describe is aimed at calculating expectation values of fields $\hat{\vphi}^a$ all at the same time, which commute among each other. However, the fact that the cyclicity of the trace may fail  would be a major obstacle in trying to extend the classical Matsubara formalism to a sort of classical real time formalism.} in the case of a classical system:
\be
\mbox{Tr} \, \hat{\rho} = \int d\vphi \, e^{-\beta H(\hat{\vphi})} \langle \vphi | \vphi \rangle \sim  \delta(0) = \infty. 
\ee

In the quantum  Matsubara approach, the trace of the density operator $\hat{\rho}=e^{-\beta H(\hat{\varphi})}$ can be written as a path integral associated with an evolution along the imaginary axis of time. This is just a consequence of the fact that the operator $H(\hat{\varphi})$, which appears in the density operator $\hat{\rho}$, is also the generator of time evolution in quantum mechanics. In the KvN approach to classical mechanics, however, the situation is different, since the generator of time evolution is no longer the Hamiltonian $H$, but the Lie derivative of the Hamiltonian flow $\mathcal{H}$ defined in equation (\ref{eq-31}). Nevertheless, we can still consider the action of the Gibbs operator $\hat{\rho}$ on the various operators defined on the KvN Hilbert space. In fact, $\hat{\rho}$ commutes with $\hat{\varphi}^a$, $\hat{c}^a$ and $\hat{\bar{c}}_a$, but it induces a change in the operators $\hat{\lambda}_a$, according to the following formula:
\be
\hat{\lambda}_a^{\prime}=\hat{\rho}^{-1} \, \hat{\lambda}_a \, \hat{\rho}=
\hat{\lambda}_a+ i \beta \, \omega_{ab}\frac{d\hat{\varphi}^b}{dt}.
\ee
From the previous equation, we can easily evaluate the change induced by the operator $\hat{\rho}$ on  the superfield $\hat{\Phi}$ defined in equation (\ref{superfield}):
\begin{eqnarray}
\hat{\rho}^{-1}\hat{\Phi}^a(t,\theta,\bar{\theta}) \, \hat{\rho} &=& \hat{\varphi}^a(t)+\theta \hat{c}^a(t)+\bar{\theta}\omega^{ab}\hat{\bar{c}}_b(t)+i\bar{\theta}\theta \omega^{ab}\hat{\lambda}_b(t)- \beta \bar{\theta}\theta \frac{d\hat{\varphi}^a}{dt} \nonumber \\
&=& \hat{\varphi}^a(t- \beta \bar{\theta} \theta)+\theta \hat{c}^a(t)+\bar{\theta}\omega^{ab}\hat{\bar{c}}_b(t)+i\bar{\theta}\theta \omega^{ab}\hat{\lambda}_b(t) \nonumber \\
&=& \hat{\Phi}^a(t- \beta \bar{\theta}\theta ,\theta, \bar{\theta}). \la{evolution}
\end{eqnarray}
According to our conventions \cite{Abrikosov:2004cf} about the Grassmann partners
of time, $\theta$ and $\bar{\theta}$,   the product $\bar{\theta}\theta$ is purely imaginary. Therefore, we can think of ``complexifying" the time variable via the following rule: 
\begin{equation}
t\, \to \, t_1+ \bar{\theta}\theta \, t_2, \label{plane} 
\end{equation}
where, for the purposes of this paper, it will be sufficient to take $t_1$ and $t_2$ as real variables\footnote{Notice however that there is a fundamental difference between the usual imaginary unit $i$ and the product $\bar{\theta}\theta$, namely that $i^2 = -1$ while $(\bar{\theta}\theta)^2=0$.}. The action of $\hat{\rho}$ on a superfield can then be interpreted, according to equation (\ref{evolution}),  as a ``time'' evolution along the $\bar{\theta}\theta$ component in the $(t_1, t_2)$ plane, as shown in Figure \ref{imaginary time evolution}. 
\begin{figure}[h!]
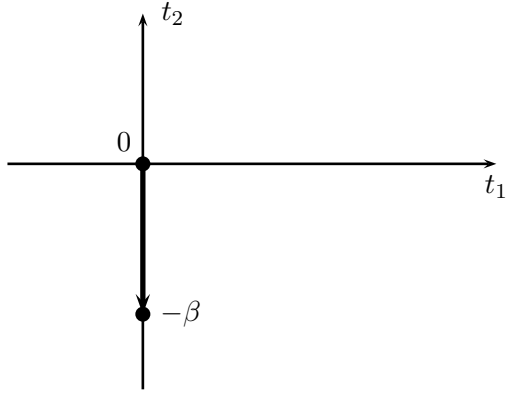

\begin{center}
\pspicture(0,3)(8,8.5)
\psset{linewidth=1pt}
\psline[]{->}(2.3,3)(2.3,8)
\psline[]{->}(0.5,6)(7,6)
\pscircle*(2.3,6){0.1}
\rput(7,5.7){$t_1$}
\rput(2.7,8){$t_2$}
\psset{linewidth=2pt}
\pscircle*(2.3,4){0.1}
\psline[]{->}(2.3,6)(2.3,4)
\rput(2.8,4){$-\beta$}
\rput(2.05,6.3){0}
\endpspicture
\end{center}
\caption{Time evolution along the $\bar{\theta}\theta$ direction.} \la{imaginary time evolution}
\end{figure}

It is also interesting to consider the action of  $\hat{\rho}$ on states in the KvN Hilbert space, and in particular on the eigenstates of $\hat{\Phi}^\phi$ in the Heisenberg picture. If we multiply the Hermitian conjugate of the eigenvalue equation
\be
\displaystyle \langle \Phi^\phi, t | \, \hat{\Phi}^\phi(t,\theta,\bar{\theta}) = \displaystyle \langle \Phi^\phi, t | \, \Phi^\phi(\theta, \bar{\theta})
\ee
by $\hat{\rho}$ and we insert the identity $\hat{\rho}\, \hat{\rho}^{-1} = \mathbb{1}$ after each bra, we easily get
\be
\displaystyle  \langle \Phi^\phi, t | \, \hat{\rho} \, \hat{\Phi}^\phi(t- \beta \bar{\theta} \theta,\theta,\bar{\theta}) =  \langle \Phi^\phi, t | \, \hat{\rho} \, \Phi^\phi(\theta,\bar{\theta}).
\ee
This equation clearly implies that $ \langle \Phi^\phi, t | \, \hat{\rho}$ is an eigenstate of the operator $\hat{\Phi}^\phi(t- \beta \bar{\theta} \theta,\theta,\bar{\theta})$, which can be indicated as follows:
\be \la{Gibbs evol 2}
\displaystyle  \langle \Phi^\phi, t | \, \hat{\rho} \equiv \langle\Phi^\phi, t- \beta \bar{\theta} \theta |.
\ee
The analogy with the quantum Matsubara formalism is striking, and equations (\ref{evolution}) and (\ref{Gibbs evol 2}) are the main results of this subsection which will allow us to set up a path integral formulation based on the statistical operator $\hat{\rho} = e^{- \beta H(\hat{\vphi})}$.

\subsection{Path integral} \la{classical matsubara path integral}

We now have all the necessary ingredients to develop a path integral formulation for the classical Matsubara approach. The first step is then to calculate the trace of $\hat{\rho}$ on the basis of the eigenstates of $\Phi^\phi$ as shown in equation (\ref{trace}) of appendix A. By virtue of equation (\ref{Gibbs evol 2}), we get:
\be
 \textrm{Tr}\, \hat{\rho}  = \int [d\Phi^{\phi}]  \, \langle \Phi^{\phi} ,0  | \, \hat{\rho} \, | \tilde{\Phi}^{\phi} , 0 \rangle =  \int [d \Phi^{\phi}] \, \langle \Phi^{\phi} , \beta \theta \bar{\theta}  | \tilde{\Phi}^{\phi} , 0 \rangle, \label{y.1}
\ee
where we have introduced a new superfield $\tilde{\Phi}^{\phi}$, which differs from the usual one $\Phi^{\phi}$ defined in (\ref{superfield}) only for the  sign in front of the Grassmann variables $c^\phi$ and $\bar{c}_\pi$:
\be \la{phitilde}
\tilde{\Phi}^{\phi} (x, \theta, \bar{\theta}) \equiv \phi (x) - \theta c^{\phi} (x) - \bar{\theta}  \bar{c}_{\pi} (x) + i \bar{\theta} \theta \lambda_{\pi}(x) = \Phi^{\phi} (x, -\theta, -\bar{\theta}).
\ee
Notice also that, for later convenience, we have inverted the order of $\theta$ and $\bar{\theta}$ in  (\ref{y.1}) compared to equation (\ref{Gibbs evol 2}). By following the standard slicing procedure, equation (\ref{y.1}) can then be rewritten in terms of a path integral. To this end, let us divide the interval $\beta \theta \bar{\theta}$ in $N+1$ subintervals of length $\epsilon\theta \bar{\theta}$ and insert $N$ times the completeness relation (\ref{a29}) for the eigenstates of $\hat{\Phi}^\phi$. The transition amplitude in (\ref{y.1}) becomes then: 
\begin{equation} \label{z.9}
\langle \Phi^{\phi}, \beta \theta \bar{\theta} | \tilde{\Phi}^{\phi} , 0 \rangle = \prod_{n=1}^{N} \int [d \Phi^{\phi}_n] \prod_{n=0}^{N}  \langle \Phi^{\phi}_{n+1} , s_{n+1} \,\theta \bar{\theta} | \Phi^{\phi}_n,s_n\theta\bar{\theta}\rangle,
\end{equation}
where $s_0\equiv 0$, $s_{N+1}\equiv \beta$, $s_{n+1}-s_n=\epsilon$ and we have chosen $\Phi^{\phi}_0=\tilde{\Phi}^{\phi}$, $\Phi^{\phi}_{N+1}\equiv \Phi^{\phi}$.
According to the definition (\ref{superfield}) of superfields, we can rewrite the Hamiltonian $H(\hat{\varphi})$ in terms of the superfields as:
\be
\displaystyle H(\hat{\varphi})=\int id\theta d\bar{\theta}\left[ i\theta \bar{\theta} H(\hat{\Phi})\right]=H(\hat{\Phi})\Bigl|_0,
\ee
where the notation $\bigl|_0$ indicates the first component of the multiplet $H(\hat{\Phi})$, which can be also obtained by setting $\theta,\bar{\theta}=0$. 
By inserting a completeness relation for the eigenstates of $\Phi^{\pi}$ in each of the infinitesimal amplitudes appearing in (\ref{z.9}) we obtain: 
\begin{eqnarray}
\langle \Phi^{\phi}_{n+1} , s_{n+1} \,\theta \bar{\theta} | \Phi^{\phi}_n,s_n\theta\bar{\theta}\rangle &=& \langle \Phi^{\phi}_{n+1} | e^{- \epsilon H (\hat{\varphi})} | \Phi^{\phi}_n \rangle \nonumber \\
& \approx &  \langle \Phi^{\phi}_{n+1} | \exp\l(- \epsilon \int d^3x \l.\fr{(\hat{\Phi}^\pi)^2}{2}\r|_0 \r)  \exp\l(- \epsilon V (\hat{\Phi}^{\phi})\bigl|_0\r)  | \Phi^{\phi}_n \rangle \nonumber \\
&=& \int [d \Phi^{\pi}_n] \, \langle \Phi^{\phi}_{n+1} | \Phi^{\pi}_n \rangle \langle \Phi^{\pi}_n |  \Phi^{\phi}_n \rangle  \, e^{- \epsilon H(\Phi_n)\bigl|_0}  \nonumber \\
&=& \int [d \Phi^{\pi}_n] \, \exp \left[  i\epsilon \int i\, d^3xd\theta d\bar{\theta}\, \l( \Phi^{\pi}_n \dot{\Phi}^{\phi}_n -  \theta \bar{\theta} \mathscr{H} (\Phi_n) \r) \right], \la{infinitesimal path integral}
\end{eqnarray}
where we have introduced the discretized time derivative $\dot{\Phi}^{\phi}_n\equiv (\Phi^{\phi}_{n+1}-\Phi^{\phi}_n)/\epsilon$ as well as the Hamiltonian density ${\mathscr H}$  for the fields associated with the Hamiltonian $H$. By combining equations (\ref{y.1}), (\ref{z.9}) and (\ref{infinitesimal path integral}) and turning to the continuum limit, we finally obtain the following path integral expression for the trace of the statistical operator $\hat{\rho}$: 
\begin{equation} \label{z.0}
\textrm{Tr}\, \hat{\rho} = \int \mathcal{D} \Phi^{\phi} \mathcal{D} \Phi^{\pi} \exp \left[ i \int_0^{\beta} ds \int i d^3xd\theta d\bar{\theta} \left( \Phi^{\pi} \dot{\Phi}^{\phi} -  \theta \bar{\theta} {\mathscr H}(\Phi) \right)\right],
\end{equation}
where the path integral is over trajectories in the space of the superfields which satisfy the boundary condition $ \Phi^{\phi} ( \beta \theta \bar{\theta},\mathbf{x}, \theta, \bar{\theta}) = \tilde{\Phi}^{\phi} ( 0,\mathbf{x},  \theta, \bar{\theta})$. The result (\ref{z.0}) can also be rewritten in a simpler way by observing that the integral over the parameter $s$ can be converted into an integral over the time variable $t(s)=\theta \bar{\theta} s$. Using the standard rules of calculus, we get:
\begin{displaymath}
\displaystyle \dot{\Phi}^{\phi}=\frac{d\Phi^{\phi}}{ds}=\theta \bar{\theta}\,\frac{d\Phi^{\phi}}{dt}, \qquad \theta\bar{\theta} ds=dt.
\end{displaymath}
The final result for the trace of the density operator is the following:
\begin{equation} \label{w.2}
\textrm{Tr} \, \hat{\rho} = \int \mathcal{D} \Phi^{\phi} \mathcal{D} \Phi^{\pi} \exp \left[ i \int_0^{-\beta \bar{\theta}\theta} d \tau \int d^3 x \left(\Phi^{\pi} \frac{d \Phi^{\phi}}{dt} -  {\mathscr H}(\Phi) \right) \right],
\end{equation}
where $d \tau = i dt  d \theta d \bar{\theta}$. This is the path integral for the Matsubara approach to classical thermal field theory. It is interesting to compare this expression with the path integral for the quantum Matsubara approach, i.e.
\begin{equation} \label{w.1}
\textrm{Tr}\,  \hat{\rho}  = \int \mathcal{D} \phi \mathcal{D} \pi \exp \left[ i \int_0^{- i \beta} d t \int d^3 x \left(\pi \frac{d \phi}{dt} -  {\mathscr H}(\varphi) \right) \right].
\end{equation}
The formal analogy between (\ref{w.2}) and (\ref{w.1}) is striking, and in fact the quantum expression can be turned into the classical one by using the two dequantization rules \cite{Abrikosov:2004cf} already mentioned at the end of section \ref{SUSY},
together with a new rule: 
\begin{description}
\item[3.] $ i\beta \,  \longrightarrow \,   \bar{\theta}  \theta \beta$.
\end{description}

\subsection{Generating functional and Feynman rules}

We can now take the path integral expression (\ref{w.2}) as our starting point to systematically derive the classical perturbative calculations in a way which closely resembles what is usually done in the quantum Matsubara formalism. In fact, we can define the generating functional as follows:
\begin{equation} \label{w.4}
\mathbf{Z} [\mathbf{J}_{\phi}] = \frac{\displaystyle \int \mathcal{D} \Phi^{\phi} \mathcal{D} \Phi^{\pi} \exp \left[ i \int_0^{-\beta \bar{\theta}\theta}  d\alpha \left( \Phi^{\pi} \frac{d \Phi^{\phi}}{dt} -  {\mathscr H}(\Phi) + \mathbf{J}_{\phi} \Phi^{\phi} \right) \right]}{\displaystyle \int \mathcal{D} \Phi^{\phi} \mathcal{D} \Phi^{\pi} \exp \left[ i \int_0^{-\beta \bar{\theta}\theta}  d\alpha \left( \Phi^{\pi} \frac{d \Phi^{\phi}}{dt} -  {\mathscr H}(\Phi) \right) \right]},
\end{equation}
where we have simplified the notation by defining $\alpha\equiv (x,\theta,\bar{\theta})$ and $d\alpha \equiv i \, d^4x \, d\theta \, d\bar{\theta}$, and by introducing the following supercurrent:
\be
\mathbf{J}_{\phi} (\alpha) \equiv - \Lambda^{\pi} (x) - i \theta \eta^{\pi} (x)+ i \bar{\theta} \bar{\eta}_{\phi} (x) - i \bar{\theta} \theta J_{\phi} (x) .
\ee
As usual, expectation values can be calculated systematically by taking functional derivatives of $\mathbf{Z} [\mathbf{J}_{\phi}]$ and  setting $\mathbf{J}_{\phi} = 0$ at the end. However, it is important to realize that, this procedure allows us to calculate only time-independent expectation values -- another point of contact with the quantum Matsubara formalism. In fact, if we parametrize  the path in the ``complexified'' time plane shown in Figure \ref{imaginary time evolution} as $t(s)=-s\bar{\theta}\theta$, then the functional derivatives w.r.t. a supercurrent evaluated on this path can be naturally defined in such a way that
\be
\displaystyle \frac{\delta {\mathbf{J}}_{\phi}(t(s),{\mathbf x}, \theta,\bar{\theta})}{\delta{\mathbf J}_{\phi}(t(s^{\prime}),{\mathbf x}^{\prime},\theta^{\prime},\bar{\theta}^{\prime})}\equiv i \delta (s-s^{\prime})\delta ({\mathbf x}-{\mathbf x}^{\prime})\delta(\theta-\theta^{\prime}) \delta(\bar{\theta}-\bar{\theta}^{\prime}).
\ee
Then, by using this equation together with the Taylor expansion $\Phi^\phi (t(s), \mathbf{x}, \theta, \bar{\theta}) = \Phi^\phi (0, \mathbf{x}, \theta, \bar{\theta}) - s \bar{\theta} \theta \, \dot{\Phi}^\phi (0, \mathbf{x}, \theta, \bar{\theta})$, we get the following result for the first derivative of the generating functional evaluated at $\mathbf{J}_{\phi} = 0$:
\be
\displaystyle \frac{1}{i}\frac{\delta {\mathbf Z}}{\delta {\mathbf J}_{\phi}(\alpha)}\Bigg|_{{\mathbf J}_{\phi}=0}= \Big\langle \frac{\delta}{\delta {\mathbf J}_{\phi}(\alpha)} \int_0^{-\beta \bar{\theta}\theta}  d \alpha' \mathbf{J}_{\phi}(\alpha') \Phi^\phi (\alpha') \Big\rangle_\beta =\theta \bar{\theta} \, \langle \phi (0, \mathbf{x})\rangle_{\beta}.
\ee
This result clearly shows that every functional derivative w.r.t. $\mathbf{J}_{\phi}(\alpha)$ will contribute with a factor of $\phi (0, \mathbf{x})$. Therefore, by using the generating functional (\ref{w.4}), we will only be able to calculate thermal correlation functions at time $t=0$. 
In fact, by taking $n$ derivatives of the generating functional w.r.t. the supercurrents and setting ${\mathbf J}_{\phi}=0$, we can get the expectation value of the product of $n$ fields $\phi$ evaluated at different \emph{space} points, but at the same time $t=0$:
\begin{equation}
\displaystyle \frac{1}{i}\frac{\delta}{\delta {\mathbf J}_{\phi}(\alpha_1)}
\cdots \frac{1}{i}\frac{\delta}{\delta {\mathbf J}_{\phi}(\alpha_n)}\,{\mathbf Z}[{\mathbf J}_{\phi}]\Bigg|_{{\mathbf J}_{\phi}=0}=\left[\prod_{i=1}^{n}\theta_i\bar{\theta}_i\right]\langle \phi (0, \mathbf{x}_1) \cdots \phi (0, \mathbf{x}_n)\rangle_{\beta}. \label{mv}
\end{equation}
A good way to see that these are indeed the only quantities we can calculate by using the generating functional (\ref{w.4}) is to realize that the coupling term between the external supercurrent ${\mathbf J}_{\phi}$ and the superfield $\Phi^\phi$ actually reduces to
\be
 i \int_0^{-\beta \bar{\theta}\theta}  d\alpha \,  \mathbf{J}_{\phi} \Phi^{\phi} = - \beta \int d^3 x \,\Lambda^\pi (0, \mathbf{x}) \phi (0, \mathbf{x}). \la{simplification}
\ee
We can now use the generating functional (\ref{w.4}) to derive a set of Feynman rules for a scalar field with quartic self-interaction described by the Hamiltonian (\ref{eq-40}). As we will see, these rules will involve only one propagator and one vertex, exactly like in the quantum Matsubara formalism. 

As usual, the propagator can be determined by neglecting the quartic interaction. To this end, it is more convenient to work in the operatorial formalism where, up to a normalization constant, the free generating functional reduces to:
\ba
{\mathbf Z}_0 [{\mathbf J}_{\phi}] &\sim& \mbox{Tr} \l[ \hat{\rho}  \, T\exp \l(  i \int_0^{-\beta \bar{\theta}\theta}  d\alpha \, \mathbf{J}_{\phi} \hat{\Phi}^{\phi}\r)\r] \nonumber \\
&\sim& \int [d \vphi] \exp \l\{ - \beta \int d^3 x \l[ \fr{\pi^2}{2} + \fr{(\nabla \phi)^2}{2} + \fr{m^2 \phi^2}{2} + \Lambda^\pi  \phi \r]_{t=0} \r\}. \la{freegen1}
\ea
Notice that, in passing from the first to the second line we have used equation (\ref{simplification}). The integral over $\pi$ in equation (\ref{freegen1}) yields an overall multiplicative constant, while integral over $\phi$ is a standard Gaussian integral that can be calculated by considering the Fourier transform of the field $\phi$ and the current $\Lambda^\pi$ w.r.t. the spatial variables only. By imposing the normalization condition ${\mathbf Z}_0 [{\mathbf J}_{\phi}=0] = 1$ one finally gets:
\ba
{\mathbf Z}_0 [{\mathbf J}_{\phi}]  &=& \exp \l\{ - \fr{\beta}{2} \int \fr{d^3 p}{(2 \pi)^3} \, \Lambda^\pi (0, \mathbf{p}) \fr{1}{E_{\mathbf{p}}^2} [\Lambda^{\pi} (0,  \mathbf{p})]^*\r\} \nonumber \\
&=& \exp \left\{- \frac{1}{2}\int_0^{-\beta  \bar{\theta} \theta} d\tau d\tau^{\prime} \int \frac{d^3p}{(2\pi)^3} \, {\mathbf J}_{\phi}(\tau,{\mathbf p})\frac{1}{\beta E_{\mathbf p}^2} {\mathbf J}_{\phi}^*(\tau^{\prime},{\mathbf p})\right\}, \la{clasmatsuprop}
\ea
where $E_{\mathbf p}=\sqrt{{\mathbf p}^2+m^2}$. Equation (\ref{clasmatsuprop}) clearly shows that the classical Matsubara formalism admits only one propagator which is given by
\be \la{matsuprop}
\parbox{20mm}{
\begin{fmfgraph}(65,30)
\fmfleft{i1}
\fmfright{o1}
\fmf{phantom}{i1,o1}
\fmffreeze
\fmf{vanilla}{i1,o1}
\fmfv{decor.shape=circle, decor.filled=full, decor.size=1.5thick}{i1,o1}
\end{fmfgraph}} \quad =
G^c_\beta (\tau, \mathbf{p}) = \fr{1}{\beta E_\mathbf{p}^2} + \mathcal{O}(\tau).
\ee
Although equation (\ref{clasmatsuprop}) clearly shows that terms of order $\mathcal{O}(\tau)$ or higher in the propagator $G^c_\beta (\tau, \mathbf{p})$ do not contribute to classical thermal correlation functions, we have included such terms in (\ref{matsuprop}) because they play a role in the comparison between $G^c_\beta (\tau, \mathbf{p})$ and the propagator of the quantum Matsubara formalism at high temperatures. In fact, according to the general result (\ref{propagator}), the propagator for the quantum Matsubara formalism in the $(t, \mathbf{p})$ space and in the high temperature limit takes the form
\be
G_\beta (t, \mathbf{p}) = \fr{n_B(E_\mathbf{p})}{2 E_\mathbf{p}} \l( e^{i E_\mathbf{p} t} + e^{\beta E_\mathbf{p}} e^{- i E_\mathbf{p} t} \r) \stackrel{\beta\to 0}{\approx} \fr{\cos(E_\mathbf{p} t)}{\beta E_\mathbf{p}^2} = \fr{1}{\beta E_\mathbf{p}^2} + \mathcal{O}(t^2).
\ee
Hence, we see that the free quantum propagator correctly reduces to the classical propagator in equation (\ref{matsuprop}) in the high temperature limit. 

For completeness, let us mention that the rule for the vertex can be easily derived by using the fact that the full generating functional can be expressed in terms of the free one as
\be
\displaystyle {\mathbf Z}[{\mathbf J}_{\phi}]=\exp \left[ -i \int_0^{- \beta \bar{\theta} \theta}d\tau d^3x \, {\mathcal V}\left(\int id\theta^{\prime}d\bar{\theta}^{\prime}\frac{\delta}{\delta {\mathbf J}_{\phi}(t,\theta^{\prime},\bar{\theta}^{\prime},{\mathbf x})}\right)\right]{\mathbf Z}_0[{\mathbf J}_{\phi}],
\ee
where ${\mathcal V}$ is the perturbation, which in general contains terms of order higher than quadratic.
In particular, in the case of a quartic interaction with coupling $g$ we obtain the following rule for the vertex in the $(\tau, \mathbf{p})$ space:
\be \la{clamatsuvertex}
\parbox{20mm}{
\begin{fmfgraph}(45,40)
\fmfleft{i1,i2}
\fmfright{o1,o2}
\fmf{vanilla}{i1,v,o2}
\fmf{vanilla}{i2,v,o1}
\fmfv{decor.shape=circle, decor.filled=full, decor.size=1.5thick}{v}
\end{fmfgraph}}
\! \! \! \! =-ig \int_{0}^{- \beta \bar{\theta} \theta} \textrm{d}\tau.
\ee

\subsection{Equivalence to the classical Thermofield Dynamics and CTP formalism}

In conclusion, it is interesting to check that the classical Matsubara formalism we developed in this section reproduces the same static results that can be obtained from the classical CTP formalism or Thermofield Dynamics. To this end, let us consider once again the one loop corrections to the two-point correlation function. Within the classical Matsubara formalism, these corrections are described by one single diagram, and according to the rules (\ref{matsuprop}) and (\ref{clamatsuvertex}) in the $(t, \mathbf{p})$ space they are equal to
\vspace{0.6cm}
\ba
\parbox{20mm}{
\begin{fmfgraph}(45,40)
\fmfleft{i1}
\fmfright{o1}
\fmf{vanilla,tension=0.4}{i1,v,v,o1}
\fmfv{decor.shape=circle, decor.filled=full, decor.size=1.5thick}{v}
\fmfv{decor.shape=circle, decor.filled=full, decor.size=1.5thick}{i1}
\fmfv{decor.shape=circle, decor.filled=full, decor.size=1.5thick}{o1}
\end{fmfgraph}} \! \! \! \! &=& -\fr{i g}{2} \int_0^{-\beta \bar{\theta} \theta} d \tau \int \fr{d^3p}{(2 \pi)^3} G_\beta^c (\tau_1 - \tau, \mathbf{k}) G_\beta^c (0, \mathbf{p}) G_\beta^c (\tau - \tau_2, \mathbf{k}) \nonumber \\
 &=& - \fr{g}{2 \beta^2 E_\mathbf{k}^4}  \int \fr{d^3p}{(2 \pi)^3} \fr{1}{E_\mathbf{p}^2} . \la{matsudiagrm}
\ea
This results should be compared with the one we already calculated in section \ref{eqivCTP} and shown in equation (\ref{1loopclasrealtime}). By taking the Fourier transform w.r.t. 0-th component of the four momentum $k_0$ and setting the time equal to zero we get
\ba
(A)+(B) &=& \int \fr{d k_0}{2 \pi} \l\{ - i \Delta_{\beta} (p) [ G_R (p) + G_A (p)] \r\} \fr{ig}{2\beta} \int \fr{d^3 p}{(2\pi)^3} \fr{1}{E_\mathbf{p}^2}  \\
&=& - \fr{g}{2 \beta^2 E_\mathbf{k}}\int d k_0 \fr{d}{d(k_0^2)} \delta(k^2-m^2) \int \fr{d^3 p}{(2\pi)^3} \fr{1}{E_\mathbf{p}^2} =  - \fr{g}{2 \beta^2 E_\mathbf{k}^4}  \int \fr{d^3p}{(2 \pi)^3} \fr{1}{E_\mathbf{p}^2} ,\nonumber
\ea
and thus we recover the same result we obtained within the classical Matsubara formalism. This very simple example is sufficient to illustrate how, in general, static results can be obtained much more easily within the classical Matsubara formalism, since fewer diagrams are needed. In our example, the single diagram (\ref{matsudiagrm}) should be contrasted with the four diagrams shown, for example, in equation (\ref{one loop diagrams}). Nevertheless, the classical Matsubara formalism has its limitations, and when it comes to calculating dynamical quantities, one is forced to adopt one of the classical approaches discussed in the previous sections.

\section{Conclusions}

In this paper we studied the connection between classical and quantum thermal field theory at high temperatures. To this end, we worked, at the classical level, within the framework of the KvN and CPI formulations of classical mechanics.

After reviewing briefly the most common formulations of quantum thermal field theory and the KvN and CPI approaches to classical mechanics, we presented the path integral approach to classical thermal field theory explored in ref. \cite{Jeon:2004dh}. We explicitly derived a set of Feynman rules to systematically calculate classical thermal correlation functions of a classical scalar field with quartic self-interaction, and we showed how the free propagators of the quantum CTP formalism are well approximated by the classical propagators at high temperatures. We also improved upon the work done in \cite{Jeon:2004dh} by showing how the cancellations occurring among classical diagrams can be seen as a natural consequence of a universal $N=2$ supersymmetry.  

In the second part of the paper, we developed two alternative approaches to classical thermal field which turned out to the related to the quantum Thermofield Dynamics and Matsubara formalism at high temperatures. In both cases, we derived the Feynman rules for a classical scalar field with quartic self-interaction, and we explicitly showed how the free quantum propagators reduce to the classical ones in the high temperature limit. 

The equivalence of these three approaches to classical thermal field theory was illustrated by considering the one loop corrections to the 2-point thermal correlation function. A more in depth study of higher order corrections and the cancellations which occur among classical diagrams due to the universal supersymmetry will be the main focus of a follow up paper to appear shortly \cite{Catta10}. In that paper, it will be shown how the use of super-diagrams in the classical Matsubara formalism greatly simplifies classical calculations. 

The fact that the classical Matsubara formalism is the only one that is entirely formulated in terms of superfields is actually puzzling, and the development of a superfield formulation for the classical TFD and CTP approaches would further strengthen the formal analogy between classical and quantum thermal field theories. This is however a topic which we leave for future study.

\acknowledgments

No thanks would be sufficient to acknowledge the generosity and the contribution of Danilo Mauro to this work. E.G. would like to thank S.Jeon and G.Aarts for useful correspondence and  E.Cattaruzza, A.Neto for discussions. This work has been supported by grants from INFN, MIUR and the University of Trieste.

\appendix

\section{The Symplectic Scalar Product} \label{app:symplectic}

We will now briefly review the definition and some basic results concerning the symplectic scalar product. For more details, we refer the reader to \cite{Deotto:2002hy}. 

Let us start by introducing some basic formulae which do not depend on the way the scalar product is implemented. For simplicity, rather than considering the full extended Hilbert space of classical mechanics, we will restrict ourselves to the subspace spanned by $\hat{c}^a$ and $\hat{\bar{c}}_a$. Because of the algebra (\ref{eq-30}) satisfied by such operators, there must exist a state $|0 -, 0 -\rangle$ such that 
\begin{equation} \label{0-0-}
\hat{c}^q  | 0 \,-, 0 \, - \rangle = \hat{c}^p  | 0 \,-, 0 \, - \rangle = 0 .
\end{equation}
By letting the operators $\hat{\bar{c}}_a$ act on this state, we can then define three more states as follows:
\begin{eqnarray}
&&| 0 \,+, 0 \, - \rangle \equiv \hat{\bar{c}}_q | 0 \,-, 0 \, - \rangle \\
&&| 0 \,-, 0 \, + \rangle \equiv - \, \hat{\bar{c}}_p | 0 \,-, 0 \, - \rangle \\
&&| 0 \,+, 0 \, +\rangle \equiv \hat{\bar{c}}_p | 0 \,+, 0 \, - \rangle = \hat{\bar{c}}_p \, \hat{\bar{c}}_q | 0 \,-, 0 \, - \rangle. \label{00others}
\end{eqnarray}
The four states introduced above form a basis for the Hilbert space on which $\hat{c}^a$ and $\hat{\bar{c}}_a$ are acting. Equation (\ref{0-0-}) shows that the state $|0 -, 0 -\rangle$ can be regarded as a simultaneous eigenstate of $\hat{c}^q$ and $\hat{c}^p$ with zero eigenvalue, but we can also consider more generic eigenstates of these two operators with Grassmann eigenvalues $\alpha^q$ and $\alpha^p$:
\begin{equation}
| \alpha^q \, -, \alpha^p \, - \rangle = e^{- \alpha^q \hat{\bar{c}}_q} \, e^{- \alpha^p \hat{\bar{c}}_p} |0 \, -, 0 \, - \rangle.
\end{equation}
Similarly, we can also consider eigenstates of the operators $\hat{\bar{c}}_a$. For example, the state 
\begin{equation} \la{eq:a6}
| \alpha^q \, -, \alpha_p \, + \rangle = e^{- \alpha^q \hat{\bar{c}}_q} \, e^{- \alpha_p \hat{c}^p} |0 \, -, 0 \, + \rangle
\end{equation}
is a simultaneous eigenstate of $\hat{c}^q$ and $\hat{\bar{c}}_p$.

Let us now consider the Hermitian conjugate of the expressions (\ref{0-0-}) through (\ref{00others}) and denote with $\langle 0 \, +, 0 - |$ the bra dual to the ket $| 0 \, +, 0 \, - \rangle$ (and similarly for the other bras). By using the Hermiticity properties of $\hat{c}^a$ and $\hat{\bar{c}}_a$ with respect to the symplectic scalar product, namely $(\hat{c}^a)^\dag = i \omega^{a b} \hat{\bar{c}}_b$, we easily obtain:
\begin{eqnarray}
&& \langle 0 \,-, 0  - |  \hat{\bar{c}}_p =  \langle 0 \,-, 0 - | \hat{\bar{c}}_q = 0  \\
&& \langle  0 \,+, 0  - | = - i \, \langle 0 \,-, 0  - | \hat{c}^p  \\
&& \langle 0 \,-, 0  + | = - i  \, \langle 0 \,-, 0  - | \hat{c}^q  \\
&& \langle 0 \,+, 0  + | = i \, \langle 0 \,+, 0  - | \hat{c}^q = \langle 0 \,-, 0 - | \hat{c}^p \hat{c}^q.
\end{eqnarray}
If we now impose the normalization condition 
\begin{equation} \label{normalization}
\langle 0 \, +, 0 + | 0 \, +, 0 \, + \rangle \equiv 1,
\end{equation}
we can calculate all the scalar products among the states which make up the basis of the Hilbert space by solely using the algebra (\ref{eq-30}) of the operators $\hat{c}^a$ and $\hat{\bar{c}}_a$ . 
It turns out that the only non-vanishing products are
\begin{eqnarray} \label{non vanishing products}
&&\langle 0 \, -, 0  + | 0 \, +, 0 \, - \rangle = i \\
&&\langle 0 \, +, 0  - | 0 \, -, 0 \, + \rangle = - i  \\
&&\langle 0 \, -, 0  - | 0 \, -, 0 \, - \rangle = - 1. \label{non definite positive}
\end{eqnarray}
Equation (\ref{non definite positive}) clearly shows that the symplectic scalar product is not positive definite. This is unfortunately an unavoidable feature of the (extended) KvN formalism: in fact, it was shown in \cite{Deotto:2002hy} that the requirement that time evolution is unitary necessarily implies a scalar product which is not positive definite, and vice versa. Nevertheless, this result does not cause any trouble at the physical level and it has an interesting physical interpretation \cite{Deotto:2002hy}. Because of equations (\ref{normalization}) through (\ref{non definite positive}),  the completeness relation reads
\begin{eqnarray}
&&  | 0 \, +, 0 \, + \rangle \langle 0 \, +, 0 + | - i \, | 0 \, +, 0 \, - \rangle \langle 0 \, -, 0 + | + \nonumber \\
&&  \quad \, + i \, | 0 \, -, 0 \, + \rangle \langle 0 \, +, 0 - | - | 0 \, -, 0 \, - \rangle \langle 0 \, -, 0 - | = \mathbb{1}
\end{eqnarray}
and, consequently, the trace of a Grassmann-even operator $\hat{A}$ is
\begin{equation}
\mbox{Tr} \hat{A} = \langle 0+,0+| \hat{A} |0+,0+\rangle -i \langle 0-,0+| \hat{A} |0+,0-\rangle + i \langle 0+,0-| \hat{A} |0-,0+\rangle -\langle 0-,0-| \hat{A} |0-,0-\rangle.
\end{equation}

By using equations (\ref{normalization}) through (\ref{non definite positive}), it is also possible to calculate the scalar product between eigenstates of $\hat{c^a}$ and $\hat{\bar{c}}_a$. In particular,  three useful results that we used in this paper are \cite{Deotto:2002hy}
\ba
\langle (i \beta_p)^* - , (i \beta^q)^* + | \alpha^q - , \alpha_p + \rangle &=& i \, \delta(\beta^q - \alpha^q) \, \delta(\beta_p - \alpha_p) \la{scalprod1}\\
\langle (- i \beta^p)^* + , (- i \beta_q)^* - | \alpha^q - , \alpha_p + \rangle &=& - i \exp \l( \beta^p \alpha_p + \beta_q \alpha^q \r)\la{scalprod2} \\
\langle ( i \beta_p)^* - , (- i \beta_q)^* - | \alpha^q - , \alpha^p - \rangle &=& - \exp \l( \beta_p \alpha^p + \beta_q \alpha^q \r) \la{scalprod3} ,
\ea
where we denoted with $\langle \alpha^q -, \alpha^p +|$ the Hermitian conjugate of the ket $| \alpha^q -, \alpha^p + \rangle$, and so on. Notice that, because of the relation $(\hat{c}^a)^\dag = i \omega^{a b} \hat{\bar{c}}_b$, the bras in equations (\ref{scalprod1}), (\ref{scalprod2}) and (\ref{scalprod3}) are eigenstates of $(\hat{c}^q, \hat{\bar{c}}_p)$, $(\hat{c}^p, \hat{\bar{c}}_q)$ and $(\hat{\bar{c}}_q, \hat{\bar{c}}_p)$ respectively, and the various factors of $i$ were introduced, together with the complex conjugate operation, in order to simplify the form of the eigenvalues.
It is also possible to derive the following completeness relations involving the eigenstates of $\hat{c}^a$ and $\hat{\bar{c}}_a$ introduced above:
\begin{eqnarray}
& \displaystyle \int  d \alpha^q d \alpha^p | \varphi, \alpha^q -, \alpha^p - \rangle \langle (-i \alpha^{p})^* +, \left(i \alpha^{q}\right)^* +, \varphi | = \mathbb{1} & \\
& \displaystyle i \int d\alpha^q d\alpha_p | \alpha^q - , \alpha_p + \rangle \langle ( i\alpha_p)^*\, -, (i\alpha^q)^* + | =\mathbb{1} , & \la{completeness1} \\
& \displaystyle i \int d\alpha_q d\alpha^p | \alpha_q + , \alpha^p - \rangle \langle ( - i\alpha^p)^* \, + ,(- i\alpha_q)^* - | =\mathbb{1} , & \la{completeness2}
\end{eqnarray}
and to use these relations to calculate the trace of a Grassmann-even operator $\hat{A}$. For example, by using equation (\ref{completeness1}) we get
\begin{eqnarray}
\mbox{Tr} \hat{A} &=& \mbox{Tr} \left\{ i \int d\alpha^q d\alpha_p | \alpha^q - , \alpha_p + \rangle \langle ( i\alpha_p)^*\, -, (i\alpha^q)^* + | \hat{A} \right\}  \nonumber \\ 
&=&  i \int d\alpha^q d\alpha_p \langle ( i\alpha_p)^*\, -, (i\alpha^q)^* + | \hat{A}  | (- \alpha^q) - , (- \alpha_p) + \rangle. \la{trace}
\end{eqnarray}
Notice that, in going from the first to the second line, the sign of the eigenvalues within the ket has changed due to the fact that we have interchanged the position of the bra and the ket. 

To conclude this appendix, let us apply the general results introduced above to the extended Hilbert space of classical mechanics. In particular, it is interesting to see how such results take a much simpler form when they are expressed in terms of superfields \cite{Abrikosov:2004cf}. In fact, according to our notation and to the definition of the superfields given in equation (\ref{superfield}), their eigenstates are
\ba
| \Phi^\phi \rangle &\equiv& | \phi, \lambda_\pi, c^\phi -, \bar{c}_\pi + \rangle \\
| \Phi^\pi \rangle &\equiv& | \lambda_\phi, \pi, \bar{c}_\phi +, c^\pi - \rangle.
\ea
Then, the eigen-bras of the superfields are defined as follows
\ba
\langle \Phi^\phi | &\equiv& \langle \phi,  \lambda_\pi, (i \bar{c}_\pi)^* -, (i c^\phi)^* + | \\
\langle \Phi^\pi | &\equiv& \langle \lambda_\phi, \pi, (- i c^\pi)^* + , (- i \bar{c}_\phi)^* -|,
\ea
and the scalar products between such eigenstates can be easily deduced from equations (\ref {scalprod1}) and (\ref {scalprod2}):
\ba
\langle \Phi^\phi_1 | \Phi^\phi_2 \rangle &=& \delta (\Phi^\phi_1 - \Phi^\phi_2) \la{a26}\\
\langle \Phi^\pi | \Phi^\phi \rangle &=& \mathcal{N}  \exp \l( \int d^3 x \, d \theta d \bar{\theta} \, \Phi^\pi \Phi^\phi\r), \la{a27}
\ea
where $\mathcal{N}$ is an irrelevant overall constant and we have introduced the functional delta
\be
\delta (\Phi^\phi_1 - \Phi^\phi_2) \equiv i \, \delta(\phi_1 - \phi_2) \delta (\lambda_{\pi \, 1} - \lambda_{\pi \, 2}) \delta (c^\phi_1 - c^\phi_2) \delta (\bar{c}_{\pi \,1} - \bar{c}_{\pi \,2}) .
\ee
Equations (\ref{completeness1}) and (\ref{completeness2}) imply that the eigenstates of $\Phi^\phi$ and $\Phi^\pi$ must satisfy the following completeness relations:
\be
\int [d \Phi^\phi] | \Phi^\phi \rangle \langle \Phi^\phi | = \int [d \Phi^\pi] | \Phi^\pi \rangle \langle \Phi^\pi | = \mathbb{1}, \la{a29}
\ee
where the integration measures are defined as $[d \Phi^\phi] \equiv i [d\phi] [d\lambda_\pi] [dc^\phi][d\bar{c}_\pi]$ and\break  $[d \Phi^\pi] \equiv i [d\pi] [d\lambda_\phi] [d\bar{c}_\phi][dc^\pi]$, and the square brackets indicate that  the integrals are carried out over field configurations at fixed time. Finally, according to equation (\ref{trace}) the trace of a Grassmann-even operator acting on the KvN extended space can be calculated on the basis of the eigenstates of $\Phi^\phi$ as
\be \la{superfieldtrace}
\mbox{Tr} \hat{A} = \int [d \Phi^\phi] \langle \Phi^\phi | \hat{A} | \tilde{\Phi}^\phi \rangle, 
\ee
where we have introduced a new superfield $\tilde{\Phi}^{\phi}$, which differs from the usual one $\Phi^{\phi}$ defined in (\ref{superfield}) only for the sign in front of the Grassmann variables $c^\phi$ and $\bar{c}_\pi$, as shown in equation (\ref{phitilde}).
Equations (\ref{a26}), (\ref{a27}), (\ref{a29}) and (\ref{superfieldtrace}) are the main results that were used in section \ref{sec:classicalMatsubara} to give a path integral formulation of the classical Matsubara formalism.


\section{Free Propagator for the Grassmann variables $c^a$ and $\bar{c}_a$} \label{app: ghost propagator}

In this appendix we will derive the Feynman rules for the propagators of $c^a$ and $\bar{c}_a$ by explicitly calculating the path-integral:
\be
Z_2 =  \mathcal{N} \int  \mathcal{D} \bar{c} \, \mathcal{D} c\, \exp \left\{ i \int d^4 x \, [ i \bar{c}_a \dot{c}^a - i \bar{c}_\phi c^\pi - i \bar{c}_\pi ( \nabla^2 - m^2 ) c^\phi - i \bar{\eta}_\phi c^\phi - i \bar{c}_\pi \eta^\pi ] \right\},
\ee
where $\mathcal{N}$ is determined by the requirement that $Z_2 = 1$ when $\eta_a = \bar{\eta}^a = 0$.  As a first step, we can rewrite the argument of the exponential in a more compact form as follows:
\begin{eqnarray}
& \displaystyle i \int d^4 x \, d^4 x^{\prime} \, \delta( x^{\prime} - x) \, [\bar{c}_a (x^{\prime}) (i \delta^a_b \partial_t - \tensor{M}{^a_b} ) c^b (x)+ \bar{c}_a (x^{\prime})\tensor{P}{^a_b} \eta^b (x) + \bar{\eta}_a (x^{\prime}) \tensor{Q}{^a_b} c^b (x) ] =& \nonumber \\
&= i [ \bar{c} \,D \, c + \bar{c} \, P \, \eta + \bar{\eta} \, Q \, c],& \label{q.7}
\end{eqnarray}
where we have defined:
$$ \tensor{M}{^a_b} \equiv \left( \begin{array}{cc} 0 & i \\ i( \nabla^2 - m^2) & 0 \end{array} \right), \qquad 
\tensor{P}{^a_b} \equiv \left( \begin{array}{cc} 0 & 0 \\ 0 & - i \end{array} \right), \qquad
 \tensor{Q}{^a_b} \equiv \left( \begin{array}{cc} -i & 0 \\ 0 & 0 \end{array} \right), $$
and 
\begin{eqnarray}
D(x^{\prime}, x) &\equiv& \delta( x^{\prime} - x) \, (i \delta^a_b \partial_t - \tensor{M}{^a_b} ) \nonumber \\ 
P(x^{\prime}, x) &\equiv& \delta( x^{\prime} - x)\, \tensor{P}{^a_b} \\
Q(x^{\prime}, x) &\equiv& \delta( x^{\prime} - x) \, \tensor{Q}{^a_b}. \nonumber
\end{eqnarray} 
According to the usual Feynman prescription, let us modify the differential operator $D$ by adding an infinitesimal parameter $\epsilon$ in such a way that $D \rightarrow D + i \epsilon$. Equation (\ref{q.7}) can now be rewritten as 
\begin{equation} \label{q.8}
i [\bar{c} + \bar{\eta} \, Q\, (D + i\epsilon)^{-1}] (D + i\epsilon) [ c + (D + i\epsilon)^{-1} \, P \, \eta] - i \bar{\eta} \, Q\, (D + i\epsilon)^{-1} \, P \, \eta.
\end{equation}
After introducing new integration variables  $\bar{c}^{\prime} \equiv \bar{c} + \bar{\eta} \, Q\, (D + i\epsilon)^{-1}$ and $c^{\prime} \equiv c + (D + i\epsilon)^{-1} \, P \, \eta$, the calculation of  the path integral becomes straightforward, and by requiring that $Z_2$ be normalized to one, we obtain the final result:
\begin{equation}
Z_2 = \exp \left( - i \bar{\eta} \, Q\, (D + i\epsilon)^{-1} \, P \, \eta \right).
\end{equation}

If we now denote with $G$ the inverse of the differential operator $(D + i\epsilon)$, the relation $(D + i\epsilon) \, G =~\mathbf{1}$ can be written more explicitly in Fourier space as follows
\begin{equation} \label{c1}
\left( \begin{array}{cc} p^0 + i \epsilon & - i \\ i E_{{\bf p}}^2 &  p^0 + i \epsilon \end{array} \right) G(p) = \left( \begin{array}{cc} 1  & 0 \\0 &  1 \end{array} \right).
\end{equation}
This last equation can be put in diagonal form via a change of basis done via the matrix:
$$ W(p) = \left( \begin{array}{cc} i/E_{{\bf p}} & - 1/2 \\ 1 & - i E_{{\bf p}}/2  \end{array} \right),  \qquad \quad \left( \mbox{det} W(p) = 1 \right).$$
In fact, defining $G_D(p) \equiv  W^{-1} (p) G(p) W(p)$, the (\ref{c1}) becomes:
$$ \left( \begin{array}{cc} p^0 + i \epsilon - E_{{\bf p}} & 0 \\ 0 &  p^0 + i \epsilon + E_{{\bf p}} \end{array} \right) G_D (p)  = \left( \begin{array}{cc} 1  & 0 \\0 &  1 \end{array} \right), $$
From this we can get the solution:
\begin{equation}
G_D (p) =  \left( \begin{array}{cc} 1/[p^0 + i \epsilon - E_{{\bf p}}] & 0 \\ 0 &  1/[p^0 + i \epsilon + E_{{\bf p}}] \end{array} \right).
\end{equation}
The Feynman rules in momentum space are obtained by calculating:
$$
- i Q(p) \, G(p) \,  P(p) = - i Q(p) \, W(p) \, G_D(p) \, W^{-1}(p) \, P(p) =  \left( \begin{array}{cc} 0 & - 1/[(p^0 + i \epsilon)^2 - E_{{\bf p}}^2] \\ 0 &  0 \end{array} \right).
$$
Notice that the only non-zero compoment is the one that connect  the currents  $\eta^p$ e $\bar{\eta}_q$. Expanding up to the first order  in $\epsilon$ the term  $(p^0 + i \epsilon)^2$ and re-defining $\epsilon \rightarrow \epsilon /2$, we get that this component is the opposite of the retarted propagator in momentum space:
$$ G_R (p) = \frac{1}{p^2 - m^2 + i\epsilon p^0}.$$


\end{fmffile}

\bibliographystyle{hunsrt}
\bibliography{biblio}

\begin{thebibliography}{10}

\bibitem{d'Enterria:2006cg}
D.~G. d'Enterria.
\newblock {High-energy heavy-ions physics: From RHIC to LHC}.
\newblock {\em Nucl. Phys.}, A782:215--223, 2007, nucl-ex/0608049.

\bibitem{Tannenbaum:2007dx}
M.~J. Tannenbaum.
\newblock {Heavy ion physics at RHIC}.
\newblock {\em Int. J. Mod. Phys.}, E17:771--801, 2008, nucl-ex/0702028.

\bibitem{QGP}
K.~Yagi, T.~Hatsuda, and Y.~Miake.
\newblock {\em Quark-gluon plasma: from Big Bang to Little Bang}, volume~23 of
  {\em Cambridge monographs on particle physics, nuclear physics and
  cosmology}.
\newblock Cambridge University Press, Cambridge, UK, 2005.
\newblock 446 p.

\bibitem{Grigoriev:1988bd}
D.~Y. Grigoriev and V.~A. Rubakov.
\newblock {Soliton pair creation at finite temperatures. Numerical Study in
  (1+1)-dimensions}.
\newblock {\em Nucl. Phys.}, B299:67--78, 1988.

\bibitem{Mueller:2002gd}
A.H. Mueller and D.T. Son.
\newblock On the equivalence between the {Boltzmann} equation and classical
  field theory at large occupation numbers.
\newblock {\em Phys. Lett.}, B582:279--287, 2004, hep-ph/0212198.

\bibitem{Stockamp:2004qu}
T.~Stockamp.
\newblock {Classical approximation of the Boltzmann equation in high energy
  QCD}.
\newblock {\em J. Phys.}, G32:39--46, 2006, hep-ph/0408206.

\bibitem{Aarts:1996qi}
G.~Aarts and J.~Smit.
\newblock Finiteness of hot classical scalar field theory and the plasmon
  damping rate.
\newblock {\em Phys. Lett.}, B393:395--402, 1997, hep-ph/9610415.

\bibitem{Aarts:1997kp}
G.~Aarts and J.~Smit.
\newblock Classical approximation for time-dependent quantum field theory:
  diagrammatic analysis for hot scalar fields.
\newblock {\em Nucl. Phys.}, B511:451--478, 1998, hep-ph/9707342.

\bibitem{Aarts:1997ve}
G.~Aarts.
\newblock {Renormalizability of hot classical field theory}, 1997,
  hep-ph/9707440.

\bibitem{Aarts:1999wj}
G.~Aarts, B.-J. Nauta, and C.~G. van Weert.
\newblock {Divergences in real-time classical field theories at non- zero
  temperature}.
\newblock {\em Phys. Rev.}, D61:105002, 2000, hep-ph/9911463.

\bibitem{Aarts:2001yx}
G.~Aarts.
\newblock {Spectral function at high temperature in the classical
  approximation}.
\newblock {\em Phys. Lett.}, B518:315--322, 2001, hep-ph/0108125.

\bibitem{Cooper:2001bd}
F.~Cooper, A.~Khare, and H.~Rose.
\newblock Classical limit of time-dependent quantum field theory: A
  {Schwinger-Dyson} approach.
\newblock {\em Phys. Lett.}, B515:463--469, 2001, hep-ph/0106113.

\bibitem{MSR}
P.~C. Martin, E.~D. Siggia, and H.~A. Rose.
\newblock {Statistical Dynamics of Classical Systems}.
\newblock {\em Phys. Rev.}, A8:423--437, 1973.

\bibitem{Jeon:2004dh}
S.~Jeon.
\newblock The {Boltzmann} equation in classical and quantum field theory.
\newblock {\em Phys. Rev.}, C72:014907, 2005, hep-ph/0412121.

\bibitem{Gozzi:1989bf}
E.~Gozzi, M.~Reuter, and W.D. Thacker.
\newblock Hidden {BRS} invariance in classical mechanics: ({II}).
\newblock {\em Phys. Rev.}, D40:3363--3377, 1989.

\bibitem{Schwinger:1960qe}
J.~S. Schwinger.
\newblock Brownian motion of a quantum oscillator.
\newblock {\em J. Math. Phys.}, 2:407--432, 1961.

\bibitem{Keldysh:1964ud}
L.~V. Keldysh.
\newblock Diagram technique for nonequilibrium processes.
\newblock {\em Zh. Eksp. Teor. Fiz.}, 47:1515--1527, 1964.

\bibitem{Matsubara:1955ws}
T.~Matsubara.
\newblock A new approach to quantum statistical mechanics.
\newblock {\em Prog. Theor. Phys.}, 14:351--378, 1955.

\bibitem{Umezawa:1982nv}
H.~Umezawa, H.~Matsumoto, and M.~Tachiki.
\newblock {\em Thermo field dynamics and condensed states}.
\newblock North-holland, Amsterdam, Netherlands, 1982.
\newblock 591 p.

\bibitem{Koopman}
B.O. Koopman.
\newblock Hamiltonian systems and transformations in {H}ilbert space.
\newblock {\em Proc. Natl. Acad. Sci.}, 17:315--318, 1931.

\bibitem{vonNeumann1}
J.~von Neumann.
\newblock Zur {O}peratorenmethode in der klassischen {M}echanik.
\newblock {\em Ann. Math.}, 33:587--642, 1932.

\bibitem{vonNeumann2}
J.~von Neumann.
\newblock Zus$\ddot{\mbox{a}}$tze zur {A}rbeit ``{Z}ur
  {O}peratorenmethode...''.
\newblock {\em Ann. Math.}, 33:789--791, 1932.

\bibitem{Abrikosov:2004cf}
A.A. Abrikosov, Jr., E.~Gozzi, and D.~Mauro.
\newblock Geometric dequantization.
\newblock {\em Ann. Phys.}, 317:24--71, 2005, quant-ph/0406028.

\bibitem{Deotto:2002hy}
E.~Deotto, E.~Gozzi, and D.~Mauro.
\newblock Hilbert space structure in classical mechanics: ({I}).
\newblock {\em J. Math. Phys.}, 44:5902--5936, 2003, quant-ph/0208046.

\bibitem{DeWitt:1992cy}
B.S. DeWitt.
\newblock {\em Supermanifolds}.
\newblock Cambridge monographs on mathematical physics. Cambridge University
  Press, Cambridge, UK, 2nd edition, 1992.
\newblock 407 p.

\bibitem{Niemi:1983nf}
A.~J. Niemi and G.~W. Semenoff.
\newblock {Finite Temperature Quantum Field Theory in Minkowski Space}.
\newblock {\em Ann. Phys.}, 152:105--129, 1984.

\bibitem{Altherr:1993tn}
T.~Altherr.
\newblock {Introduction to thermal field theory}.
\newblock {\em Int. J. Mod. Phys.}, A8:5605--5628, 1993, hep-ph/9307277.

\bibitem{Das:1997gg}
A.K. Das.
\newblock {\em Finite temperature field theory}.
\newblock World Scientific, Singapore, 1997.
\newblock 404 p.

\bibitem{LeBellac:1996}
M.~Le~Bellac.
\newblock {\em Thermal Field Theory}.
\newblock Cambridge University Press, Cambridge, UK, 1996.
\newblock 270~p.

\bibitem{Kubo:1957mj}
R.~Kubo.
\newblock Statistical mechanical theory of irreversible processes. 1. {G}eneral
  theory and simple applications in magnetic and conduction problems.
\newblock {\em J. Phys. Soc. Jap.}, 12:570--586, 1957.

\bibitem{Martin:1959jp}
P.C. Martin and J.S. Schwinger.
\newblock Theory of many particle systems. {I}.
\newblock {\em Phys. Rev.}, 115:1342--1373, 1959.

\bibitem{Bernard:74}
C.~W. Bernard.
\newblock {Feynman Rules for Gauge Theories at Finite Temperature}.
\newblock {\em Phys. Rev.}, D9:3312--3320, 1974.

\bibitem{Jackiw:74}
L.~Dolan and R.~Jackiw.
\newblock {Symmetry Behavior at Finite Temperature}.
\newblock {\em Phys. Rev.}, D9:3320--3341, 1974.

\bibitem{Carta:2005fq}
P.~Carta, E.~Gozzi, and D.~Mauro.
\newblock {Koopman-von Neumann formulation of classical Yang-Mills theories:
  I}.
\newblock {\em Annalen Phys.}, 15:177--215, 2006, hep-th/0508244.

\bibitem{DaniloPhD}
D.~Mauro.
\newblock {Topics in Koopman-von Neumann Theory}.
\newblock {\it PhD Thesis}, quant-ph/0301172.

\bibitem{Marsden}
R.~Abraham and J.E. Marsden.
\newblock {\em Foundations of Mechanics}.
\newblock Addison Wesley, 2nd edition, 1978.
\newblock 806~p.

\bibitem{Kurchan03}
S.~Tanase-Nicola and J.~Kurchan.
\newblock Statistical mechanical formulation of lyaupunov exponents.
\newblock {\em J.Phys.A: Math.Gen.}, 36:10299--10334, 2003, cond-mat.0210380.

\bibitem{Gozzi:1999uq}
E.~Gozzi and M.~Regini.
\newblock Addenda and corrections to work done on the path-integral approach to
  classical mechanics.
\newblock {\em Phys. Rev.}, D62:067702, 2000, hep-th/9903136.

\bibitem{Gozzi:1989xz}
E.~Gozzi and M.~Reuter.
\newblock Algebraic characterization of ergodicity.
\newblock {\em Phys. Lett.}, B233:383--392, 1989.

\bibitem{Penco:2006wn}
R.~Penco and D.~Mauro.
\newblock {Perturbation theory via Feynman diagrams in classical mechanics}.
\newblock {\em Eur. J. Phys.}, 27:1241--1250, 2006, hep-th/0605061.

\bibitem{Catta10}
E.~Cattaruzza, E.~Gozzi, and A.~Neto.
\newblock Diagrammar in classical scalar field theory.
\newblock {\em To appear}, 2010.

\bibitem{Nakazato:1990kk}
H.~Nakazato, K.~Okano, L.~Schulke, and Y.~Yamanaka.
\newblock {Symmetries in stochastic quantization and Ito-Stratonovich related
  interpretation}.
\newblock {\em Nucl. Phys.}, B346:611--631, 1990.

\bibitem{Ezawa:1984bm}
H.~Ezawa and J.~R. Klauder.
\newblock {Fermion without fermions: the Nicolai map revisited}.
\newblock {\em Prog. Theor. Phys.}, 74:904--915, 1985.

\bibitem{Sato:1976hy}
M.-a. Sato.
\newblock {Operator Ordering and Perturbation Expansion in the Path Integration
  Formalism}.
\newblock {\em Prog. Theor. Phys.}, 58:1262--1270, 1977.

\end{thebibliography}

\end{document}